\documentclass{aa}  

\usepackage{graphicx}
\usepackage{txfonts}
\usepackage{siunitx}

\usepackage{natbib,twoopt}
\usepackage[breaklinks=true]{hyperref} %
\bibpunct{(}{)}{;}{a}{}{,} %
\newcommandtwoopt{\citeads}[3][][]{\href{http://adsabs.harvard.edu/abs/#3}{\def\hyper@linkstart##1##2{}\let\hyper@linkend\@empty\citealp[#1][#2]{#3}}}
\newcommandtwoopt{\citepads}[3][][]{\href{http://adsabs.harvard.edu/abs/#3}%
{\def\hyper@linkstart##1##2{}%
\let\hyper@linkend\@empty\citep[#1][#2]{#3}}}
\newcommandtwoopt{\citetads}[3][][]{\href{http://adsabs.harvard.edu/abs/#3}%
{\def\hyper@linkstart##1##2{}%
\let\hyper@linkend\@empty\citet[#1][#2]{#3}}}
\newcommandtwoopt{\citeyearads}[3][][]%
{\href{http://adsabs.harvard.edu/abs/#3}
{\def\hyper@linkstart##1##2{}%
\let\hyper@linkend\@empty\citeyear[#1][#2]{#3}}}
\makeatother

\usepackage{color}
\definecolor{mygreen}{RGB}{0,128,0}

\hypersetup{colorlinks=true,linkcolor=blue,citecolor=blue,urlcolor=blue}

\usepackage[normalem]{ulem}

\begin{document}

   \title{The Baade-Wesselink projection factor of RR Lyrae stars}
   \subtitle{Calibration from OHP/SOPHIE spectroscopy\thanks{Based in part on observations made at Observatoire de Haute Provence (CNRS), France} and \textit{Gaia} DR3 parallaxes}

   \author{Garance Bras\inst{1} \and Pierre Kervella\inst{1} \and Boris Trahin\inst{2} \and Piotr Wielgórski\inst{3} \and Bartłomiej Zgirski\inst{4} \and Antoine Mérand\inst{5} \and Nicolas Nardetto\inst{6} \and Alexandre Gallenne\inst{4,7} \and Vincent Hocdé\inst{3} \and Louise Breuval\inst{8} \and Anton Afanasiev\inst{1} \and Grzegorz Pietrzyński\inst{3} \and Wolfgang Gieren\inst{4}
          }

   \institute{LESIA, Observatoire de Paris, Université PSL, CNRS, Sorbonne Université, Université Paris Cité, 5 place Jules Janssen, 92195 Meudon, France\\
              \email{garance.bras@obspm.fr}
              \and Space Telescope Science Institute, 3700 San Martin Drive, Baltimore, MD 21218, USA \and Nicolaus Copernicus Astronomical Center, Polish Academy of Sciences, Bartycka 18, 00-716 Warszawa, Poland \and Universidad de Concepción, Departamento de Astronomía, Casilla 160-C, Concepción, Chile \and European Southern Observatory, D-85748 Garching, Munich, Germany \and Université Côte d'Azur, Observatoire de la Côte d'Azur, CNRS, Laboratoire Lagrange, France \and French-Chilean Laboratory for Astronomy, IRL 3386, CNRS, Casilla 36-D, Santiago, Chile \and 
              Department of Physics and Astronomy, Johns Hopkins University, Baltimore, MD 21218, USA}

   \date{}

  \abstract
   {The application of the parallax of pulsation (PoP) technique to determine the distances of pulsating stars implies the use of a scaling parameter, namely the projection factor ($p$-factor), which is required to transform disc-integrated radial velocities (RVs) into photospheric expansion velocities. The value of this parameter is poorly known and is still debated.
   Most present applications of the PoP technique assume a constant $p$-factor. However, it may actually depend on the physical parameters of each star, as past studies designed to calibrate the $p$-factor (predominantly for Cepheids) led to a broad range of individual values.}
   {We aim to calibrate the $p$-factors of a sample of RR Lyrae stars (RRLs) in order to compare them with classical Cepheids (CCs). Due to their higher surface gravity, RRLs have more compact atmospheres, and therefore provide a valuable comparison with their supergiant siblings.}
   {We determined the $p$-factor of 17 RRLs by modelling their pulsation using the SPIPS code. The models are constrained using \textit{Gaia} DR3 parallaxes, photometry, and new RVs that we collected with the OHP/SOPHIE spectrograph.
   We carefully examine the different steps of the PoP technique, in particular the method used to determine the RV from spectra using the classical approach based on the cross-correlation function (CCF).}
   {The method employed to extract the RV from the CCF has a strong impact on the $p$-factor, that is, of up to 10\%. However, this choice of method results in a global scaling of the $p$-factor, and only marginally affects the scatter of $p$ within the sample for a given method.
   Over our RRL sample, we find a mean value of $p = 1.248 \pm 0.022$ for RVs derived using a Gaussian fit of the CCF.
   There is no evidence for a different value of the $p$-factor of RRLs,
   although its distribution for RRLs appears significantly less scattered ($\sigma \approx 7\%$) than that for CCs ($\sigma \approx 12\%$).}
   {The $p$-factor does not appear to depend in a simple way on fundamental stellar parameters (pulsation period, radius, metallicity, amplitude of the RV curve).
   We argue that large-amplitude dynamical phenomena occurring in the atmospheres of RRLs (and CCs) during their pulsation affect the relative velocity of the spectral line-forming regions compared to the velocity of the photosphere.
   }

   \keywords{}

   \maketitle

\section{Introduction}

Pulsating stars, and in particular Cepheids and RR Lyrae, are essential indicators for the determination of distances in the local Universe. Precise calibration of the relations between their apparent luminosity, period of pulsation \citep{Leavitt1912,Breuval2022}, and metallicity is therefore particularly important.
These calibrations are based on geometrical determinations of the distance using, for instance, parallaxes \citep[see][for example]{Groenewegen2018} or detached eclipsing binaries \citep{Riess2019,Pietrzynski2019DEB}.
Such empirical relations play a central role in the extragalactic distance ladder. Period--luminosity relations of Cepheids in the Milky Way, the Magellanic Clouds, and NGC 4258 are used to calibrate the luminosity of type Ia supernovae, whose distances enable us to determine the Hubble constant $H_0$ \citep{SH0ES2022}.
However, this empirical measurement of $H_0$ differs by $5\sigma$ from the prediction inferred from the Planck cosmological microwave background data assuming the $\Lambda$CDM model \citep{Planck2020}.
In this context, there is growing interest in better understanding the physics of pulsating stars in order to explore possible biases that  their period--luminosity relations  may incur on the distance ladder. 

The parallax-of-pulsation (PoP) technique to determine distances of pulsating stars, also known as the Baade-Wesselink technique \citep{Baade1926, Wesselink1946}, was first introduced by \cite{Lindemann1918}.
It is based on a comparison of the variation of the linear radius of a pulsating star to the variation of its angular diameter, the ratio of these two quantities giving the distance of the star.
The variation of photospheric radius is estimated from the integration of the pulsation velocity curve of the photosphere of the star over its pulsation cycle.
The resulting radius curve is then compared to the angular diameter curve, which can be either  measured directly by interferometry for nearby stars, or predicted using the surface-brightness--colour relation \citep[SBCR; see][for instance]{KervellaSBCR2004,Nardetto2023}. The SBCR is generally calibrated using interferometric measurements.
Unfortunately, it is not possible to directly measure the pulsational velocity by spectroscopy, but only the radial velocity (RV), which is an averaged quantity over the stellar disc.
The ratio $p = v_\mathrm{puls}/v_\mathrm{rad}$ between the pulsation velocity $v_\mathrm{puls}$ and the RV $v_\mathrm{rad}$ is called the projection factor, or $p$-factor. Considering a uniformly bright pulsating sphere, the projection on the line of sight of the spherical distribution of the pulsation velocity vector corresponds to $p=1.5$.
The $p$-factor also incorporates the effect of limb darkening, which results in a higher weight attributed to the RV in the centre of the star compared to the limb. This results in a lower value of $p$ than the purely geometrical value, and typically around $p\sim1.3$.
In addition, $p$ is also affected by velocity gradients that may exist in the atmosphere of the star between the line-forming regions and the continuum-forming photosphere. This aspect was studied theroretically in \cite{Nardetto2004} and with observations in \cite{Nardetto2007}.
Finally, the $p$-factor also depends on the method used to estimate the RV from the spectra. Such measurements need to be taken carefully, and well documented, as demonstrated by \cite{AndersonConf2020} and \cite{BorgnietRVCep}, but this is difficult because of the asymmetry of the lines in the spectrum of a pulsating star.

The intrinsic astrophysical complexity of the $p$-factor makes its calibration challenging. Using independent distance measurements (e.g. using \textit{Gaia} parallaxes), several authors determined the $p$-factor of pulsating stars  empirically \citep{Storm2011,Ngeow2012, Breitfelder2016,Kervella2017,Nardetto2017,TrahinpCep,Nardetto2023}. 
The resulting large scatter of the determined $p$-factors led to a withdrawal of the PoP technique as a way to determine the distances of pulsating stars at the percent accuracy level.
However, studying the $p$-factor is a powerful way to progress in our understanding of the dynamics of the atmosphere of pulsating stars.
Among classical pulsators, Cepheids are known to exhibit relatively complex atmospheric dynamics \citep{Fokin1996,Hocde2020b}, including mass loss \citep{Neilson2008} and circumstellar envelopes \citep{Hocde2020,Gallenne2021}.
Conversely, RR Lyrae stars (RRLs), which present intrinsic differences in their atmospheres when compared to Cepheids (e.g. they are physically smaller than Cepheids, with a higher surface gravity, a more compact atmosphere, and a higher temperature), make for a valuable comparison in order to characterise their $p$-factor. This is the motivation behind the present study, which is intended to expand the investigation into classical Cepheids  presented by \citet{TrahinpCep} towards the lower-mass RRLs.

In Sect.~\ref{data}, we present the new RV data that we collected with the OHP/SOPHIE spectrograph, as well as the sources of ancillary literature data (photometry and effective temperature). In Sect.~\ref{RV}, we detail the methodology we used to extract RV measurements from our SOPHIE spectra. Section~\ref{SPIPS} is dedicated to the analysis of the pulsation of 17 RRLs using the SPIPS modelling tool. In Sect.~\ref{results}, we summarise our results on the $p$-factor, as well as those on the physical properties of the modelled stars. Finally, we conclude this study in Sect.~\ref{ccl}.

\section{Observational data}\label{data}

\begin{table*}
\caption{Selection of 23 RRLs in the northern hemisphere observed with the SOPHIE spectrograph at OHP, with their mean magnitude in band V ($m_V$), exposure time ($\Delta t$), and number of spectra acquired.}             
\label{table:1}      
\centering          
\begin{tabular}{c S c c c c c l l l}     
\hline\hline       
Name & $m_V$ & $\Delta t$ & $N$ & Maximum & $N_\mathrm{rebound}$ & $<S/N>$ & Refs & Refs & Refs\\
 &  & (s) & & gap & & & RV & $T_\mathrm{eff}$ & photometry \\
\hline
V0341 Aql$^\dagger$ & 10.89 & 900 & 13 & 0.17 & 2 & 22.0 & & & 35 \\
X Ari$^\dagger$ & 8.97 & 600 & 25 & 0.27 & 3 & 26.0 & 6$^\bigstar$ & 12,13 & 21$^\bigstar$,24$^\bigstar$,30$^\bigstar$,33,35 \\
RS Boo$^{\dagger,B}$ & 9.73 & 600 & 11 & 0.42 & 3 & 26.0 & 8$^\bigstar$ & 12$^\bigstar$,13,16 & 26$^\bigstar$ \\
RR Cet$^\ast$ & 9.15 & 600 & 10 & 0.22 & 3 & 25.0 & 9$^\bigstar$,10 & 12,14,16 & 28$^\bigstar$,29,30$^\bigstar$,33 \\
W CVn$^\ddagger$ & 11.18 & 900 & 19 & 0.08 & 4 & 25.0 & 5 & 12 & 23,27,35 \\
UY Cyg$^\ddagger$ & 11.10 & 900 & 16 & 0.19 & 7 & 17.0 &  & & 23$^\bigstar$,27$^\bigstar$,30$^\bigstar$,35 \\
DX Del$^\ast$ & 9.52 & 600 & 6 & 0.67 & 3 & 31.0 & 1,10$^\bigstar$ & 12$^\bigstar$,14$^\bigstar$,15,16$^\bigstar$ & 17$^\bigstar$,29,30$^\bigstar$,32,34$^\bigstar$ \\
SU Dra$^\ddagger$ & 9.27 & 600 & 19 & 0.18 & 8 & 29.0 & 9,11 & 13 & 28$^\bigstar$,30$^\bigstar$ \\
SW Dra$^\ddagger$ & 10.55 & 600 & 19 & 0.12 & 7 & 22.0 & 2,7 & 12$^\bigstar$ & 19$^\bigstar$,25$^\bigstar$,30$^\bigstar$ \\
RX Eri$^\ddagger$ & 9.22 & 600 & 18 & 0.12 & 7 & 25.0 & 9 & 12$^\bigstar$,14$^\bigstar$ & 28$^\bigstar$,30$^\bigstar$,33,35 \\
SV Eri & 9.96 & 900 & 4 & 0.62 & 0 & 24.0 &  & 12 & 23,30,33,35 \\
VX Her$^{\ast,B}$ & 11.31 & 900 & 7 & 0.56 & 4 & 28.0 &  & 12,14,16 & 23,27,30,35 \\
WZ Hya$^\ast$ & 10.27 & 900 & 9 & 0.43 & 3 & 16.0 &  & & 23 \\
RR Leo$^\ddagger$ & 9.94 & 900 & 25 & 0.11 & 9 & 22.0 & 9 & 12,14 & 28,30,35 \\
SS Leo$^\dagger$ & 11.00 & 900 & 12 & 0.45 & 6 & 19.0 & 4$^\bigstar$ & & 22$^\bigstar$,35 \\
TT Lyn$^\ddagger$ & 9.51 & 600 & 24 & 0.09 & 7 & 33.0 & 1,9 & 12 & 17,18,28$^\bigstar$,30$^\bigstar$,35 \\
AV Peg$^\ddagger$ & 9.87 & 600 & 22 & 0.13 & 7 & 21.0 & 9 & 12$^\bigstar$,14 & 28$^\bigstar$,30$^\bigstar$,31,35 \\
CG Peg & 11.13 & 900 & 2 & 0.92 & 0 & 18.0 &  & & \\
VY Ser$^\dagger$ & 9.73 & 600 & 17 & 0.18 & 4 & 24.0 & 3$^\bigstar$,11 & 13,15$^\bigstar$,16 & 20,22$^\bigstar$,26,35,36$^\bigstar$ \\
AB UMa$^\ddagger$ & 10.88 & 900 & 18 & 0.09 & 4 & 26.0 &  & & 27,30$^\bigstar$ \\
TU UMa$^\ddagger$ & 9.26 & 600 & 24 & 0.09 & 7 & 32.0 & 1,9 & 12$^\bigstar$ & 17$^\bigstar$,18$^\bigstar$,28$^\bigstar$,30$^\bigstar$,33,35 \\
UU Vir$^\dagger$ & 9.90 & 900 & 10 & 0.24 & 3 & 22.0 & 1,8,9 & 12,13 & 17,18,26,28,30,35 \\
BN Vul$^\ddagger$ & 11.13 & 900 & 23 & 0.10 & 6 & 24.0 & &  & 23,35 \\
\hline                  
\end{tabular}
\tablefoot{The maximum gap and $N_\mathrm{rebound}$ are two criteria that we used to select RV curves that are sufficiently well covered.
The three last columns list the references for the RV, effective temperature, and photometric data.
Star symbols $^\bigstar$ indicate the data sets that were effectively used in the model fitting. $^\ddagger$ refers to complete RV curves, $^\dagger$ to satisfactorily covered RV curves, and $^\ast$ to incomplete RV curves. $^B$ indicates the stars that exhibit the Blazhko effect. The absence of any symbol indicates that there are too few observations.}
\tablebib{
(1) \citet{BarnesBW1988}; (2) \citet{CacciariBW1987}; (3) \citet{CLBWVYSer1984}; (4) \citet{FernleySSLeoVYSer1990}; (5) \citet{FernleyPhotRV1993}; (6) \citet{JonesBWXAri1987}; (7) \citet{JonesBWSWDra1987}; (8) \citet{JonesBW1988}; (9) \citet{LJBW1989}; 
(10) \citet{MeylanBW1986}; (11) \citet{VarsavskyPhotRV1960}; (12) \citet{AndTeff1}; (13) \citet{AndTeff2}; (14) \citet{AndTeff3}; (15) \citet{Takeda2006}; (16) \citet{Takeda2022};
(17) \citet{BarnesBW1988}; (18) \citet{BarnesPhot1992}; (19) \citet{CacciariBW1987}; (20) \citet{CLBWVYSer1984}; (21) \citet{FernleyBWXAri1989}; (22) \citet{FernleySSLeoVYSer1990}; (23) \citet{FernleyPhotRV1993}; (24) \citet{JonesBWXAri1987}; (25) \citet{JonesBWSWDra1987}; (26) \citet{JonesBW1988}; (27) \citet{Layden2019}; (28) \citet{LJBW1989}; (29) \citet{MeylanBW1986}; (30) \citet{MonsonPhotRRL}; (31) \citet{Paczynski1965}; (32) \citet{Paczynski1966}; (33) \citet{PrestonPaczynski1964}; (34) \citet{SkillenDXDel1989}; (35) \citet{Sturch1966}; (36) \citet{VarsavskyPhotRV1960}}
\\

\end{table*}
 
\subsection{Spectroscopic data}
We focus our study on a selection of 23 fundamental RRLs observable from the Northern hemisphere; these are listed in Table~\ref{table:1}.
We obtained new spectra for these RRLs using the SOPHIE spectrograph installed at the 193cm telescope of the Observatoire de Haute-Provence in southern France. 
SOPHIE is a cross-dispersed échelle spectrograph \citep{SOPHIE} with excellent stability. This instrument typically provides a  signal-to-noise ratio (S/N) of 40 within 10\,minutes on the brightest RRLs in our sample. This relatively short exposure time is important, as 10 minutes already corresponds to 1-2\% of the pulsation cycle of a RRL with a period of 10 hours. Longer exposure times would potentially result in smearing of the RV curve and underestimation of its amplitude.
Our observations were divided into four runs of four or five nights to survey our RRL sample. We used the high resolution mode of SOPHIE, with the slow readout mode to limit the noise. 
During the night, we organised the sequence of stars according to the previous observations and the predicted phase of the star to obtain regular and complete coverage of the pulsation cycle.
We were particularly careful to properly sample the phases between 0.8 and 1 when the stars rebound at their minimal radius, as the velocity changes quickly. For these phases, we obtained several consecutive exposures when possible.
The automatic pipeline at OHP reduces the raw spectra and computes a cross-correlation function (CCF) with a specific mask. The RV is then determined as the mean value of a Gaussian profile adjusted to the CCF.
For our RRLs, we initially chose an F0 type mask suitable for their typical effective temperature.
In the present work, we recomputed the CCF using various methods that are detailed in Sect.~\ref{RV}.
Within the list of 26 stars originally selected, we obtained a complete phase coverage for 11 of them (the number of observations is greater than 15, the maximum gap in phase between two points is lower than 0.2, and the number of observations at the rebound is more than four). Six additional stars have a satisfactory phase coverage, although some non-critical phases are incompletely covered. For four stars, part of the RV curve is missing, and two stars have too few RV measurements for an independent analysis. Three stars were not observed at all and are not presented in the table. 
Finally, the two stars \object{RS Boo} and \object{VX Her} showed a Blazhko effect during the observations (as confirmed by the literature) and were removed from our sample. We also note the binarity of TU UMa \citep{Liska2016}, which is remarkable when comparing different RV datasets.
For the stars with insufficient phase coverage, we completed our RV measurements with data from the literature. A limitation of this approach is that these additional RV measurements are potentially not determined from the spectra using the same method.

\subsection{Photometric data}
Photometric data were taken exclusively from the literature. To perform the modelling as described below, a large number of observations is required, and in different bands and photometric filters, in order to constrain the pulsating star model.
Photometric light curves from \textit{Gaia} \citep{GaiaMission, GaiaDR3} and Hipparcos \citep{Hipparcos} were employed, as well as 2MASS magnitudes \citep{2MASS}.
We also used the collection of photometric data assembled by \cite{MonsonPhotRRL}, and we added photometric measurements from the following sources: 
\citet{BarnesBW1988}, \citet{CacciariBW1987}, \citet{CacciariBWSWAndSWDraSSFor1989}, \citet{CLBWVYSer1984}, \citet{FernleyPhotRV1993}, 
\citet{FernleyBWXAri1989}, \citet{FernleySSLeoVYSer1990}, \citet{JonesBWXAri1987}, \citet{JonesBWSWDra1987}, \citet{JonesBW1988}, \citet{LJBW1989}, 
\citet{LJAbsMag1990}, \citet{MeylanBW1986}, \citet{SkillenDXDel1989}, \citet{SkillenWYAntRVOctBBPup1993}, and \citet{SkillenBW1993}.
Photometry in the bluer filters ($U$, $B$, and \textit{Gaia} $B_P$ bands) is generally not included in the model fit, as the stellar atmosphere models that we are using are insufficiently accurate for comparison with observations in this wavelength range because of some effects of the microturbulence \citep{CasagrandeVandenBerg2014}.
The \textit{Gaia} $G$ band (G\_GAIA\_GAIA3 in the models) covers a very broad wavelength range, which leads to difficulties for atmosphere models and for reddening corrections. For this reason, the $G$ band light curves are not systematically used to constrain the atmosphere models along the pulsation cycle.
Generally speaking, light curves in visible bands are much more common than in infrared filters in the literature. However, as it provides a strong constraint on the effective temperature and is also mildly sensitive to interstellar reddening, infrared photometry provides a particularly valuable constraint. This encouraged us to only consider part of the visible photometric measurements available in the literature in order to give them comparable weight in the fit to infrared measurements. The choice of data sets was based on phase coverage, agreement with other data sets, and dispersion or uncertainties. The data that were effectively used for the modelling of each RRL of our sample are marked with a star symbol  $^\bigstar$  in Table~\ref{table:1}.

\subsection{Effective temperature measurements}
Spectroscopic effective temperature estimates offer a valuable constraint on the atmosphere models of the star. This is due to the fact that these estimates are independent of the interstellar reddening of the star, and therefore help to mitigate the effects of the strong correlation that exists between the effective temperature and the colour excess $E(B-V)$ of the star. 
We took such measurements from \cite{AndTeff1, AndTeff2, AndTeff3}, who used the Spectroscopy Made Easy tool, which compares a given spectrum with synthetic ones with different atmospheric parameters.
Measurements from \cite{Takeda2006} and \citet{Takeda2022} using the same approach were also used.
APOGEE \citep{APOGEE} and LAMOST \citep{LAMOST} catalogues are of little use for this kind of work as they do not provide time-series of effective temperature. APOGEE temperatures are measured from a mean spectrum and LAMOST do not provide precise epochs of measurements. However, we were able to check that these measurements are compatible with the obtained models for effective temperature.
The number of collected effective temperature estimates is small for a single star. In order to balance the different observables in the fit of the model, we set a similar weight to each kind of observable (RVs, effective temperatures, magnitudes in the visible, and magnitudes in the infrared) in the global $\chi^2$.
Even if photometric measurements are much more numerous and precise than RVs, the weight of the latter is properly balanced in the fit. A drawback of this choice is that the normalisation can potentially give disproportionate weight to the temperature measurements compared to the other observables.
When temperature measurements were found to hamper the convergence of the model, leading to phase mismatching or unlikely light-curve shapes, we omitted them from the fitting process. However, models are still consistent with the measured temperatures.

\section{Methods to determine RV and its impact on the $p$-factor}\label{RV}
One key ingredient in the procedure to compute the $p$-factor of a pulsating star is the RV measurement. From the spectra acquired to the extraction of an RV value, some steps need to be carried out carefully. The most common way to do this is by fitting a Gaussian profile to the CCF.
However, this implies a few choices. Firstly, the choice to compute the CCF with a specific mask or template. Secondly, there exists different definitions of the RV given the CCF;
it can be defined as the centroid of the line profile as well as the mean of a Gaussian fit or as a parameter of another fitted function. Such differences in the measurements have a large impact on the $p$-factor as the angular diameter curve is compared to the integrated RV curve.
One consequence is that RVs from different methods can lead to very different $p$-factors, which need to be quantified to identify the origin of the scatter of the $p$-factors in the pulsating-star population.
In this section, we  aim to quantitatively describe the impact of each of the aforementioned choices on the $p$-factor and to motivate one or the other choice, considering physical aspects but also the diversity observed in literature measurements.

\begin{figure}
   \centering
   \includegraphics[width=\hsize]{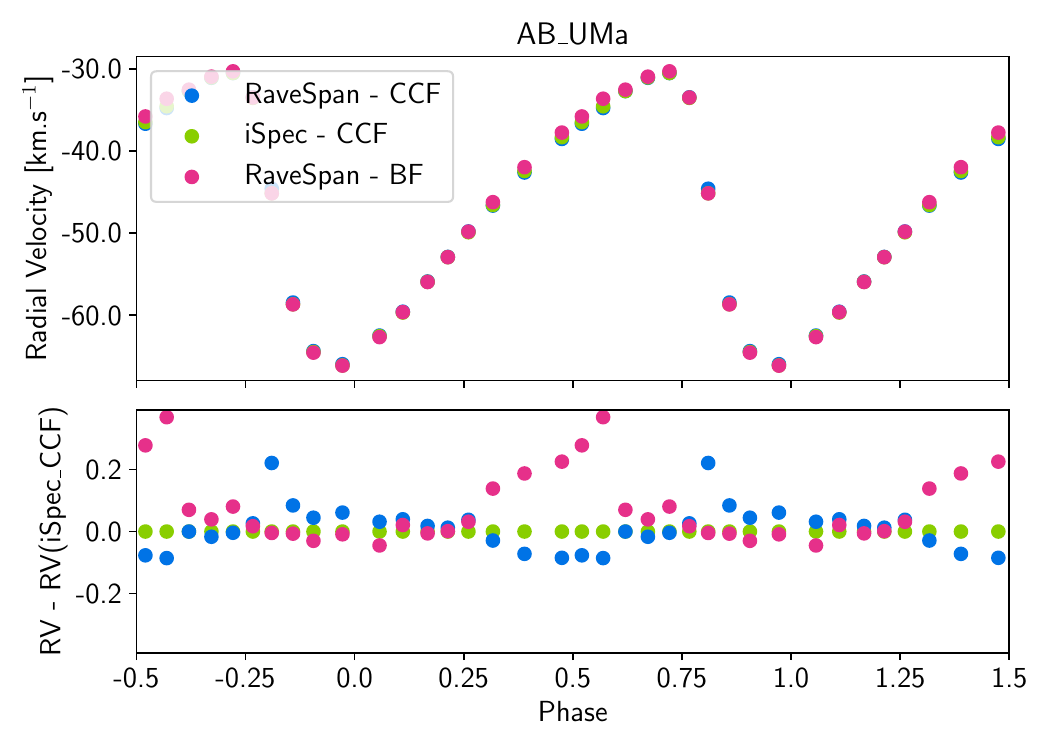}
   \caption{Comparison between the iSpec and RaveSpan codes, and between the CCF and the BF approaches in analysing the RV curve for AB UMa.}
   \label{codes_comparison_RV}
\end{figure}

\begin{figure}
   \centering
   \includegraphics[width=\hsize]{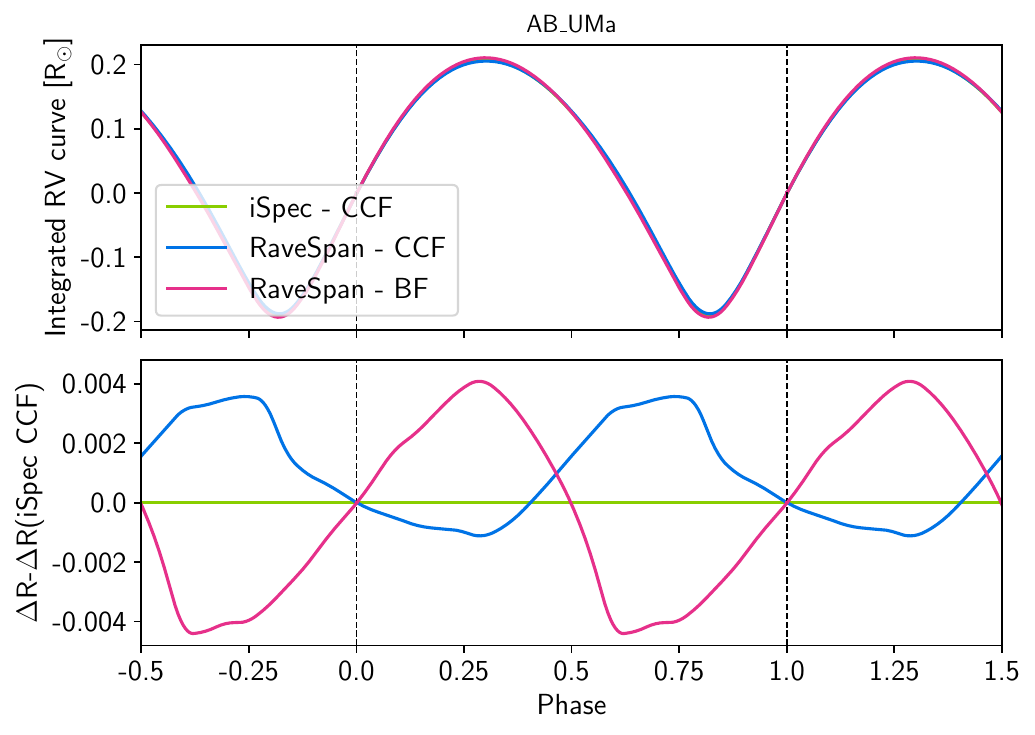}
   \caption{Comparison between the iSpec and RaveSpan codes, and between the CCF and the BF approaches in analysing the integrated RV curves from Fig. \ref{codes_comparison_RV} for AB UMa.}
   \label{codes_comparison_intRV}
\end{figure}

To compare the various methods, we interpolate the RV curve  using simple linear functions in order to avoid an impact of the choice of function used for the interpolation. This analysis is therefore only possible for really good phase coverage of the curve. We can then integrate $\Delta V = V - V_{mean}$ with time on a pulsation cycle. This results in the integrated RV curve, of which we can measure the amplitude. This amplitude is directly correlated to the $p$-factor as it corresponds to the scaling of the integrated RV curve to the angular diameter curve. For each aspect studied in the following part, we present an example using a star with good phase coverage (especially at the rebound), along with its RV and integrated RV curves. A comparison  with a reference is provided for the integrated RV curve, which is the method used afterwards.
In this work, we used two different tools in addition to the SOPHIE pipeline. Firstly, iSpec \citep{iSpec, iSpec2} was used to compute the CCF with different masks and templates, either standard or newly defined. In employing the CCF calculation method of this code, several strategies were tried, detailed in the subsequent section of the present paper.
In parallel, the RaveSpan code \citep{RaveSpan} was also used because it allows a comparison of the CCF and broadening function (BF) approaches. Both functions show similar results but the BF has the advantage that it is easily computable from a poor S/N spectrum. 
Comparisons of those different tools and of the BF and CCF approaches are presented in Figs. \ref{codes_comparison_RV} and \ref{codes_comparison_intRV} using a Gaussian profile fitting. The cross-correlation was done with the same template as that in \cite{TemplatesCoelho2005}, which is commonly used in RaveSpan, with  $T_{\mathrm{eff}}=6000 \mathrm{K}$, $\log g=3.0,$ and $\mathrm{[Fe/H]}=-1.0$. We compare the CCF and BF approaches using RaveSpan with the CCF computed by iSpec. The difference observed in the integrated RV curve is characterised by a difference of $1\%$ and $2\%$ in the amplitude respectively for the CCF and BF.

\subsection{Masks and templates}
\begin{figure}
   \centering
   \includegraphics[width=\hsize]{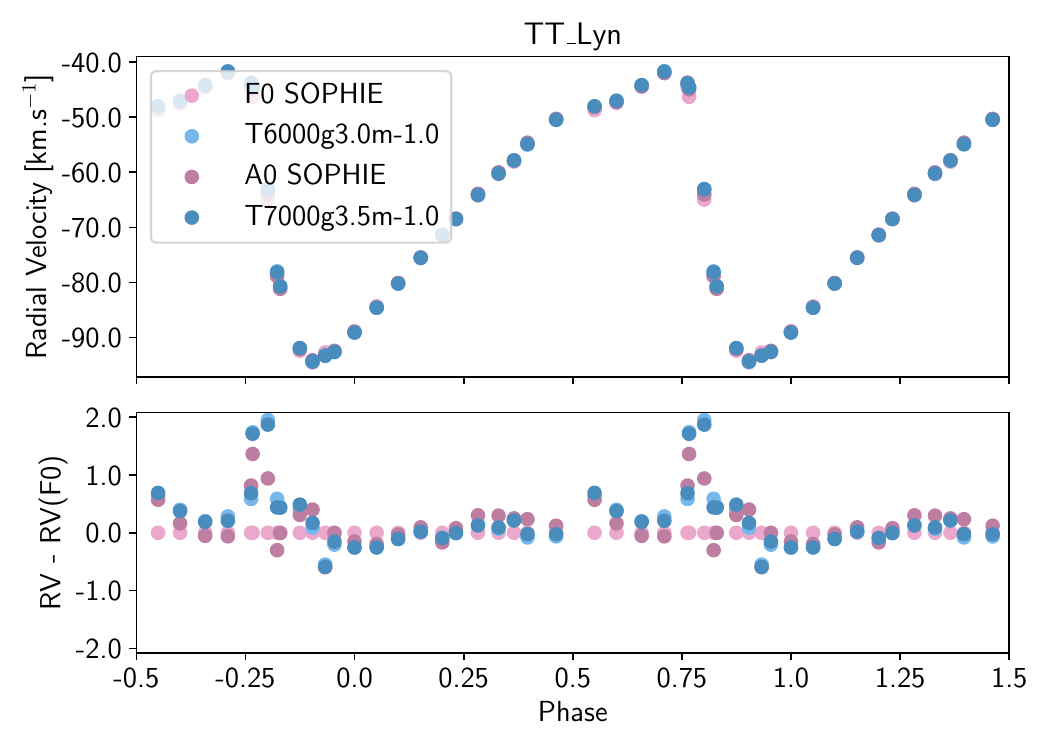}
   \caption{Comparison between masks and templates in the RV curve for TT Lyn.}
   \label{mask_comparison_RV}
\end{figure}

\begin{figure}
   \centering
   \includegraphics[width=\hsize]{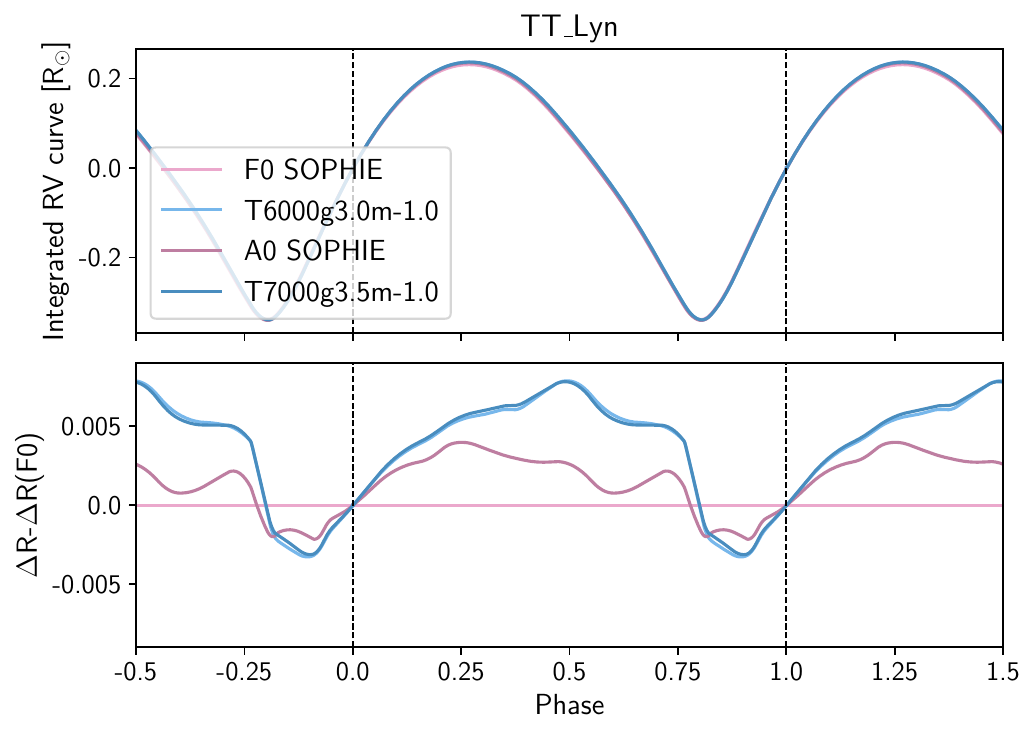}
   \caption{Comparison between masks and templates in the integrated RV curves from Fig.\ref{mask_comparison_RV} for TT Lyn. The bottom plot shows the difference between each integrated RV curve and that derived from the F0 mask.}
   \label{mask_comparison_intRV}
\end{figure}

We studied the choice between binary masks and synthetic templates for the CCF computation. We compared the standard SOPHIE masks provided in the iSpec tools, and different templates derived from synthetic spectra.
The comparison is not provided for spectral masks that are very far from the expected spectral type of RRLs, nor for templates with very different atmospheric properties. One can see the corresponding RV and integrated RV curves in Figs. \ref{mask_comparison_RV} and \ref{mask_comparison_intRV}, between two masks and two templates. Radial velocities were computed from a Gaussian fitting of the cross-correlation with an F0 and an A0 mask and with synthetic templates from \cite{TemplatesCoelho2005} with properties similar to the masks. The corresponding error in the amplitude of the integrated RV curve is of $1\%$, and we conclude that it does not have a major impact on the $p$-factor. 

Consequently, we have to keep in mind that the use of generic binary masks or templates, which contain a large number of spectral lines as they are  usually used for SOPHIE observations, will probably average the gradient in RV in the atmosphere of the star, leading to a smoothed RV curve.

The choice between masks and templates leads us to the question of the line selection. The CCF is a profile computed from a large number of different spectral lines. However, it has been shown that not all lines are pertinent for the Baade-Wesselink method \citep{Petterson2005}. Indeed, different lines from different elements are forming at different levels on the atmosphere and do not offer a good comparison to angular diameter observations. We therefore studied the impact of the selection of the lines on the $p$-factor.

\subsection{Line selection}
\begin{figure*}
   \centering
   \includegraphics[width=\hsize]{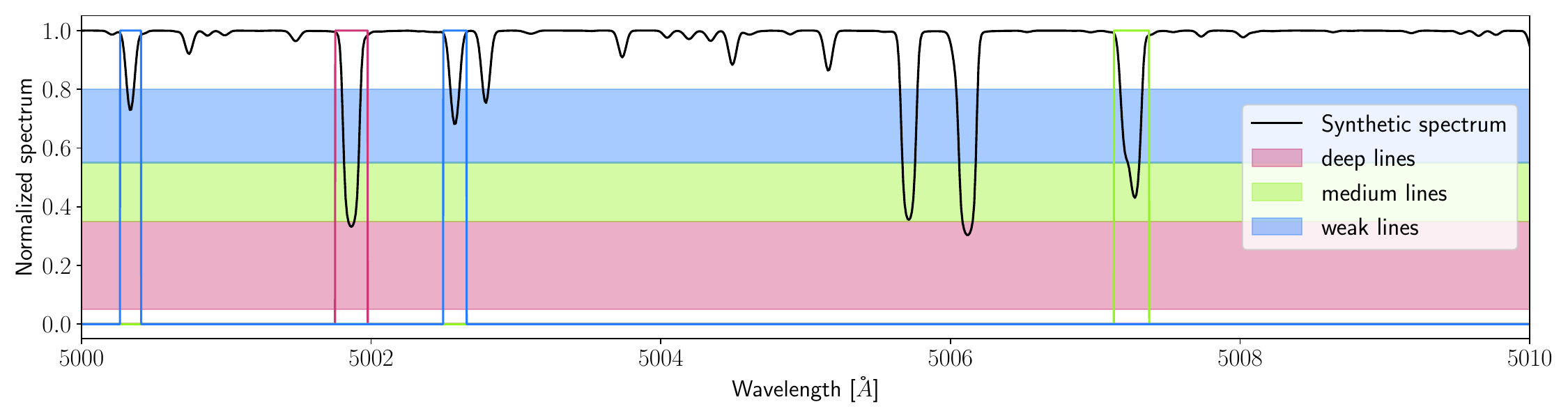}
   \caption{Portion of the synthetic spectrum used for the template construction and the corresponding built binary masks. The colour encodes the three different depth ranges, both for the background colours that define the depth region of each mask and for the plot of the binary masks themselves. 
}
   \label{masks_depths}
\end{figure*}

\begin{figure}
   \centering
   \includegraphics[width=\hsize]{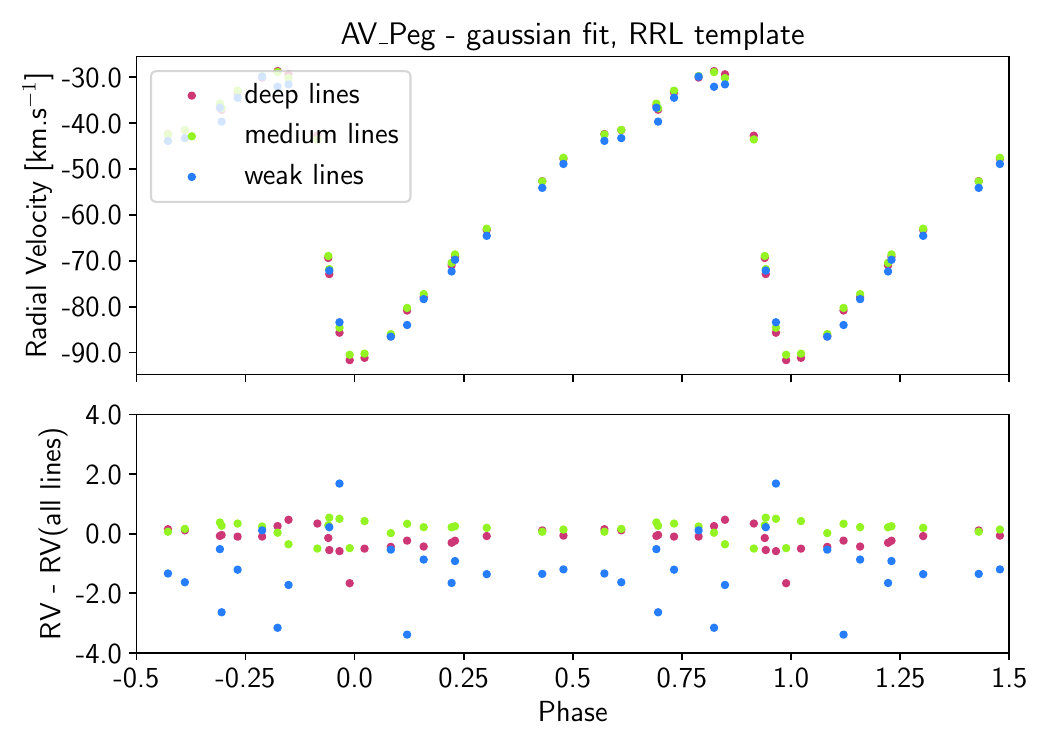}
   \caption{Comparison between weak, medium, deep, and all lines. This figure shows the RV curve from the cross-correlation of the spectra with a binary mask built for different line selection of AV Peg. The bottom plot shows the difference with the mask keeping all lines with a weight associated to its depth.}
   \label{depth_comparison_RV}
\end{figure}

\begin{figure}
   \centering
   \includegraphics[width=\hsize]{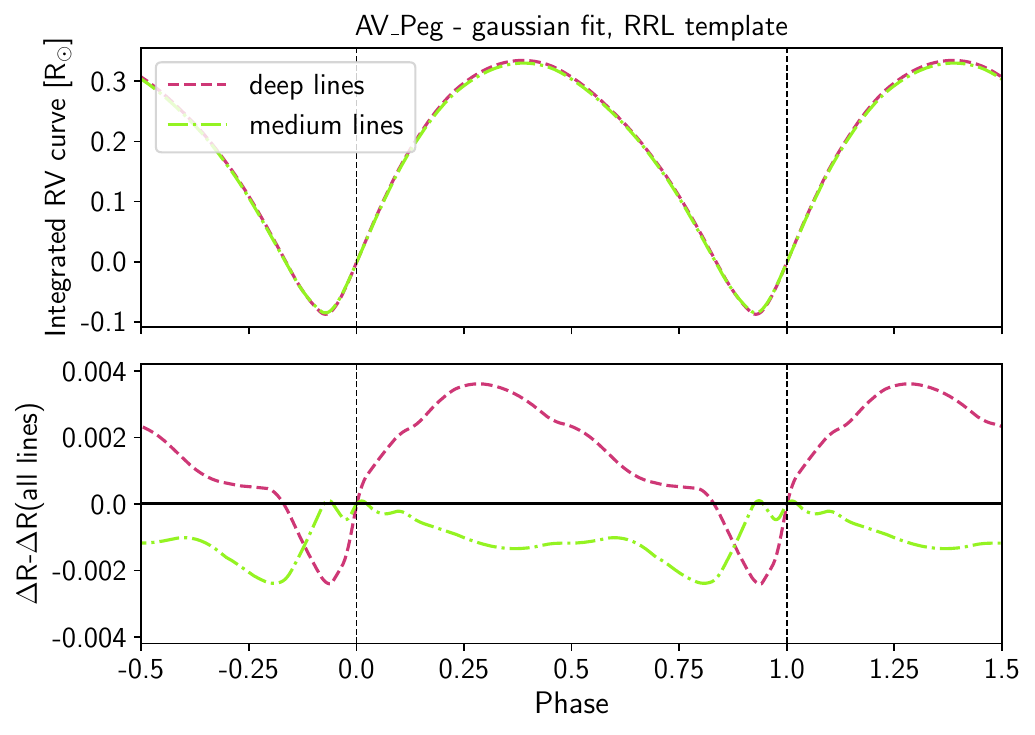}
   \caption{Comparison between medium, deep, and all lines. This figure shows the integrated RV curve (showed in Fig. \ref{depth_comparison_RV}) from Fig.~\ref{depth_comparison_RV} for AV Peg. The bottom plot shows the difference with the mask keeping all lines with a weight associated to its depth.}
   \label{depth_comparison_intRV}
\end{figure}

The CCF is created by analyzing a set of specific lines from a spectrum. These lines are determined by the mask or template used to do the cross-correlation.
\cite{Nardetto2007} reported that the amplitude of the RV curve increases with the depth of the line used to measure the RV. \cite{BorgnietRVCep} observed the effect of cross-correlating lines that are forming at different levels near the photosphere of Cepheids.
To do so, these latter authors built masks that select either weak, medium, or deep lines, considering the depths of lines to be a proxy for the optical depth. They built their masks by selecting lines in a synthetic spectrum at 5250K and with solar metallicity.
We followed the same approach for our RRLs, with new masks using a synthetic spectrum with a metallicity of -1 dex, a surface gravity of $\log g=3.0,$ and a temperature of 6200K in order to be closer to RRLs.
From such a normalised synthetic spectrum computed with the PHOENIX code \citep{Phoenix}, we applied some masks to extract the position of lines with specific ranges of depths. We defined three ranges: the deep lines for which the line depth is between 0.65 and 0.95; medium lines, with the depth between 0.45 and 0.65; and weak lines with a depth of higher than 0.2 and lower than 0.45.
We tried to extend the limits from the definition in \cite{BorgnietRVCep} for the weak lines mask to include more lines in order to improve the S/N of the CCF. In addition, a mask and a template including all these lines are also constructed. Contrary to the masks, the template is not binary, as it is weighted using the depth of each line in the synthetic spectrum. For this reason it is quite similar to the mask for deep lines as these lines have a major impact on the CCF for the general mask.
All these masks are constructed in the wavelength range from 4500$\AA$ to 6800$\AA$\footnote{Tables containing the list of wavelengths of the selected lines for all masks are available at the CDS.}. We note that the difference between the mask and the template for all lines is so small that it is not really pertinent to present both for each measurement. For each step in the analysis, we specify whether we use the binary mask or the template.
Figure \ref{masks_depths} shows part of the synthetic spectrum as an example, with the delimitation of the different ranges of depths used and the resulting masks. It is important to note that in the line selection, we only keep unblended lines in order to have a shape that is as accurate as possible for the CCF, and not smoothed by some blended lines.
The properties of the masks are summarised in Table~\ref{tab:mask_carac}.
\cite{Nardetto2007} proposed that the  CCF of the weak lines is the most pertinent for the PoP technique as it measures the velocity of the lines that were formed near the photosphere (those probed by interferometric and photometric measurements).
Unfortunately, the S/Ns of some of our observations were not sufficient to have a well-defined peak in the CCF, some of them being too noisy.
A satisfactory comparison for RV curves for different depths is therefore only provided for deep and medium lines in Fig.~\ref{depth_comparison_intRV}, as for weak lines, the phase coverage was too poor to obtain a good interpolation on the RV curve. The RV curves for the three depth-restricted masks are shown in Fig.~\ref{depth_comparison_RV}. All RVs are measured by fitting a Gaussian profile to the CCF.

\begin{figure*}
        \centering
        \includegraphics[width=\hsize]{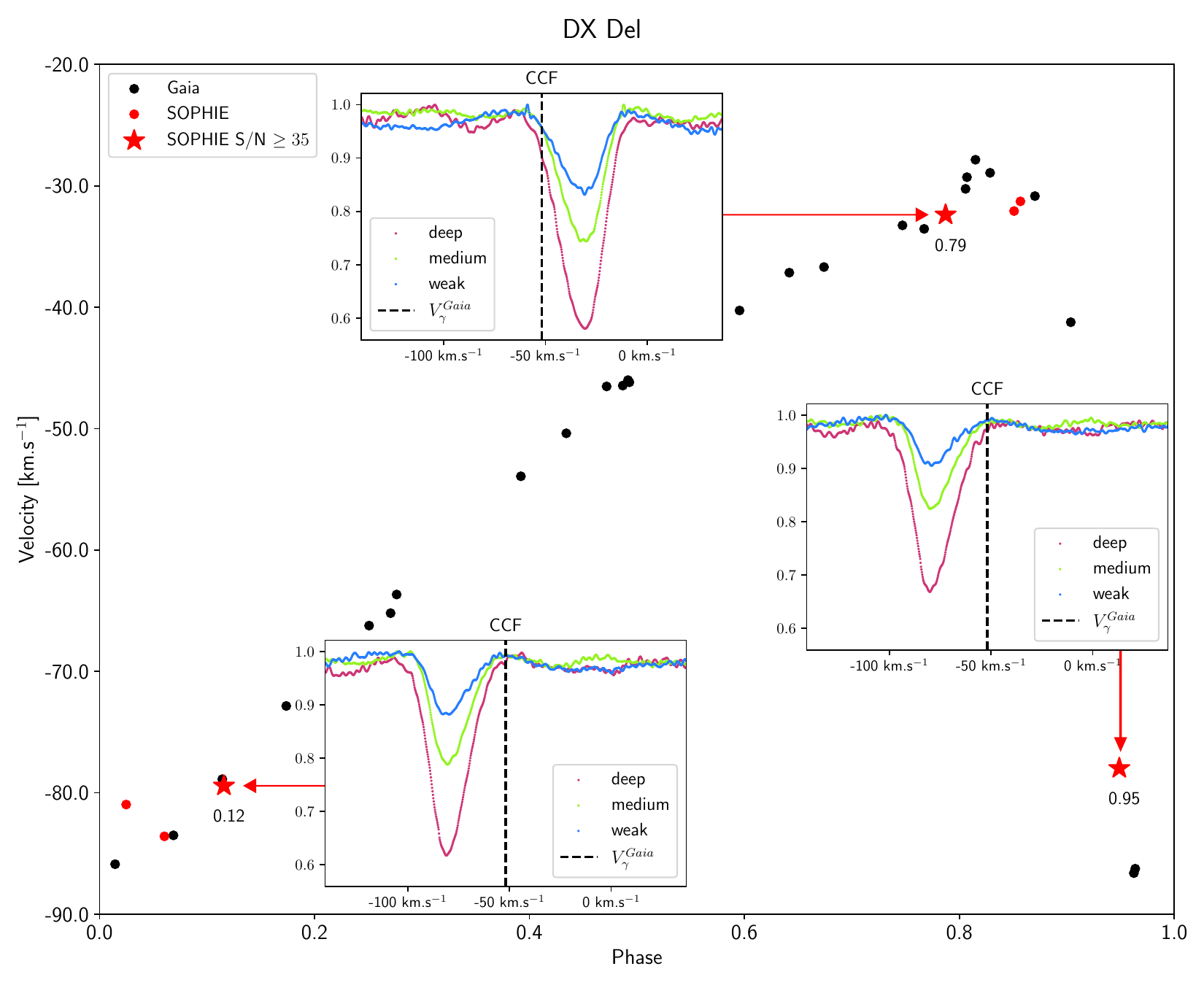}
        \caption{Evolution of the asymmetry of the CCF during the pulsation cycle of DX Del. The \textit{Gaia} RV curve for this star is shown in black for global visualisation. The red markers represent the measurements from SOPHIE and the red stars locate three spectra with a high S/N where we can compute the CCF using the three masks with different depths, shown in the inserts. The dashed line shows the mean RV extracted from the \textit{Gaia} catalogue.}
        \label{CCF_depths}
\end{figure*}

We do not see a significant difference between RV measurements from masks on different line depths ($1.3\%$ and $0.3\%$ in the amplitude of the integrated RV curve for deep lines and medium lines, respectively, with respect to all lines), which is similar to the findings of \cite{BorgnietRVCep} (see Fig.~\ref{depth_comparison_intRV}). 
Also, \cite{Petterson2005} and \cite{Nardetto2007} showed that, for Cepheids, the difference in RV for different individual lines at different layers is stronger for long-period stars.
We can extrapolate these observations and consider this effect
to be negligible for RRLs. To further investigate the different CCFs using these different masks, we chose one star with high-quality spectra (S/N typically greater than 35 at 550nm) and compared the CCF using the masks for the three ranges of depth.
Among our 23 stars, DX Del provides good spectra for this analysis at different phases, and so we can monitor how the shape evolves during the pulsation. For this star, the asymmetry of the CCF, which reverses between the phases of higher and lower velocity, is visible to the naked eye (Fig.~\ref{CCF_depths}).
This observation is compatible with the conclusions of \cite{Nardetto2006} based on single lines, who pointed out that the asymmetry of the line variation curve has, in the case of Cepheids and at first order, a shape similar to that of the RV curve itself. 
Some characterisation of the CCF is given in Table~\ref{tab:DX_Del_CCF_carac}. We computed the bisector of the line, spanning over 3$\sigma$ (with $\sigma$ being the standard deviation of the Gaussian) on both sides of the centre of the line to avoid the noise from the wings. From this bisector, we obtain $\langle b \rangle,$ the mean value of the bisector in this part of the line, and the amplitude in velocity $\Delta b$. The CCF is also fitted by a bi-Gaussian profile with widths of $\sigma_1$ and $\sigma_2$  on the bluer and redder side (positive RV) of the CCF, respectively. 
This table shows the correlation between the RV measurement from a Gaussian fit and the mean of the bisector $b$. The absolute value $\Delta b$ and the signed value $\sigma_2^{2g}-\sigma_1^{2g}$ are markers of the asymmetry of the CCF. The sign of the latter indicates which of the CCF wings is more extended:
if this parameter is positive or negative, the larger wing extension is on the red or blue part of the CCF, respectively.
Both evolve in the same way with the phase or the depth of the mask. We note that the asymmetry is systematically higher for medium and weak lines than for deep lines, which is consistent with previous observations \citep{Anderson2016}. However, it is either of the same order or higher for medium lines with respect to weak lines. \cite{BorgnietRVCep} observed that weak lines are more sensitive to the asymmetry.

\begin{table}
\caption{Characterisation of the masks. We give the number of selected lines $N$, the mean depth $\langle d \rangle$, the mean wavelength $\langle \lambda \rangle,$ and the mean width $\langle \sigma \rangle$.}             %
\label{tab:mask_carac}      %
\centering                          %
\begin{tabular}{c c c c}        %
\hline\hline                 %
 & deep lines  & medium lines & weak lines \\    %
\hline                        %
   $N$ & 80 & 155 & 140\\
   $\langle d \rangle$ & 0.72 & 0.55 & 0.37\\
   $\langle \lambda \rangle$ (nm) & 498.7 & 524.7 & 539.0\\
   $\langle \sigma \rangle$ (nm) & 0.029 & 0.020 & 0.017\\
   
\hline                                   %
\end{tabular}
\end{table}

\begin{table}
\caption{RV and asymmetry measurements on three spectra of DX Del. All velocities are in km.s$^{-1}$. $Q$ is the quality factor as defined by \cite{BorgnietRVCep}. Uncertainties smaller than 0.01 km.s$^{-1}$ in RV are not shown.}
\label{tab:DX_Del_CCF_carac}
\centering
\begin{tabular}{c | c | c c c}
\hline\hline
 & & deep & medium & weak\\
 \hline
 & RV$_g$ & -79.71 & -79.11 & -79.50$_{\pm0.01}$ \\
 & Q & 6.6 & 6.5 & 5.8 \\
$\phi = 0.12$ & $\langle b \rangle$  & -79.63 & -78.85 & -79.04 \\
 & $\Delta b$  & 2.6 & 4.2 & 3.5 \\
 & $\sigma_2^{2g}-\sigma_1^{2g}$ & 2.2 & 3.9 & 3.8\\
 \hline\hline
  & RV$_g$ & -32.35 & -32.33 & -32.69$_{\pm0.01}$ \\
 & Q & 6.2 & 6.1 & 5.7 \\
$\phi = 0.79$ & $\langle b \rangle$ & -32.47 & -32.54 & -32.93 \\
 & $\Delta b$ & 3.5 & 4.5 & 3.9\\
 & $\sigma_2^{2g}-\sigma_1^{2g}$ & -2.8  & -3.9 & -3.7\\
\hline\hline
 & RV$_g$ & -78.24 & -77.67 & -78.11$_{\pm0.01}$ \\
 & Q & 6.6 & 6.6 & 6.0 \\
$\phi = 0.95$ & $\langle b \rangle$ & -78.11 & -77.33 & -77.68 \\
 & $\Delta b$ & 4.2 & 4.9 & 3.6 \\
 & $\sigma_2^{2g}-\sigma_1^{2g}$ & 3.4 & 4.8 & 4.0\\
 
 \hline
\end{tabular}
\end{table}

\subsection{Measurement method on the CCF}
The RV is then measured through the Doppler shift of the CCF. However, spectral lines in pulsating stars show a high level of asymmetry, which can bias the various possible estimators of this Doppler shift to a varying extent.
For instance, a Gaussian fit of the CCF may be biased due to the asymmetry. An alternate approach is to fit a bi-Gaussian profile, defined as a Gaussian with different widths for the two sides of the profile.
The bi-Gaussian profile reproduces the CCF of most of the data  well (and therefore the line asymmetry derived from it will be properly determined), but does not produce a suitable RV for the projection factor (besides the additional information of the asymmetry that it provides, its mean value is almost consistent with the velocity associated to the minimum of the line, albeit less noisy) because it is highly sensitive to the line asymmetry generated, for example, by rotation. Instead, the centroid method, which is insensitive to rotation and width variation, is often preferred \citep{Nardetto2006}. 

\begin{figure}
   \centering
   \includegraphics[width=\hsize]{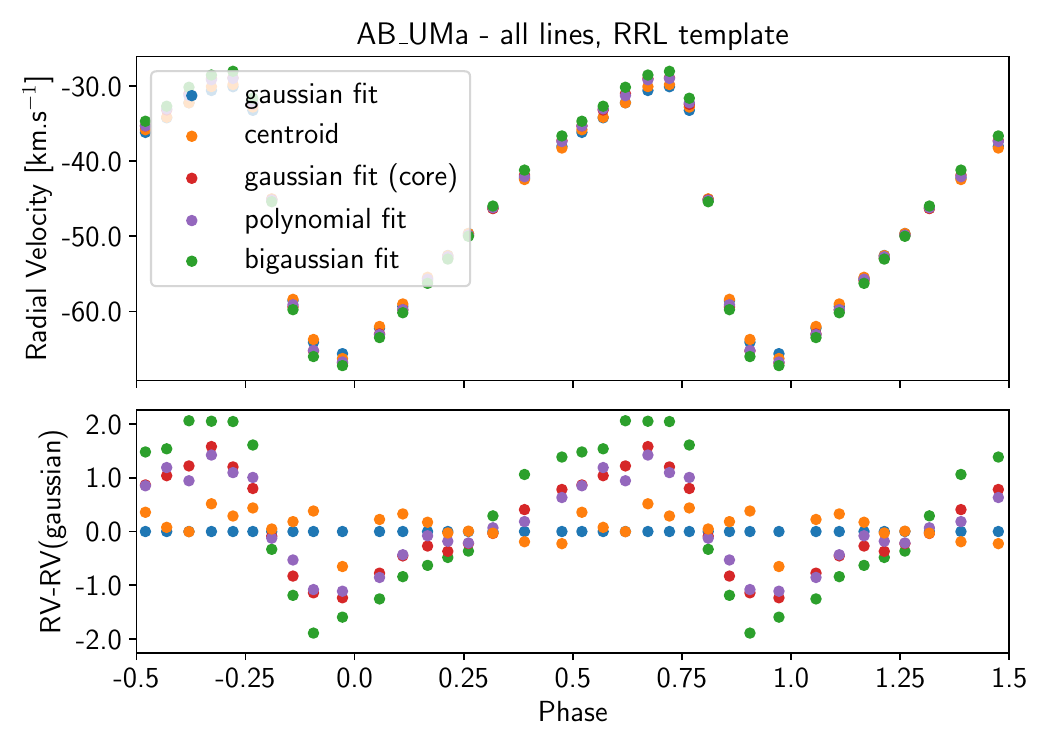}
   \caption{Comparison between definitions of RV given a CCF. The cross-correlation is computed with all lines weighted by depth mask. 
   }
   \label{def_comparison_RV}
\end{figure}

\begin{figure}
   \centering
   \includegraphics[width=\hsize]{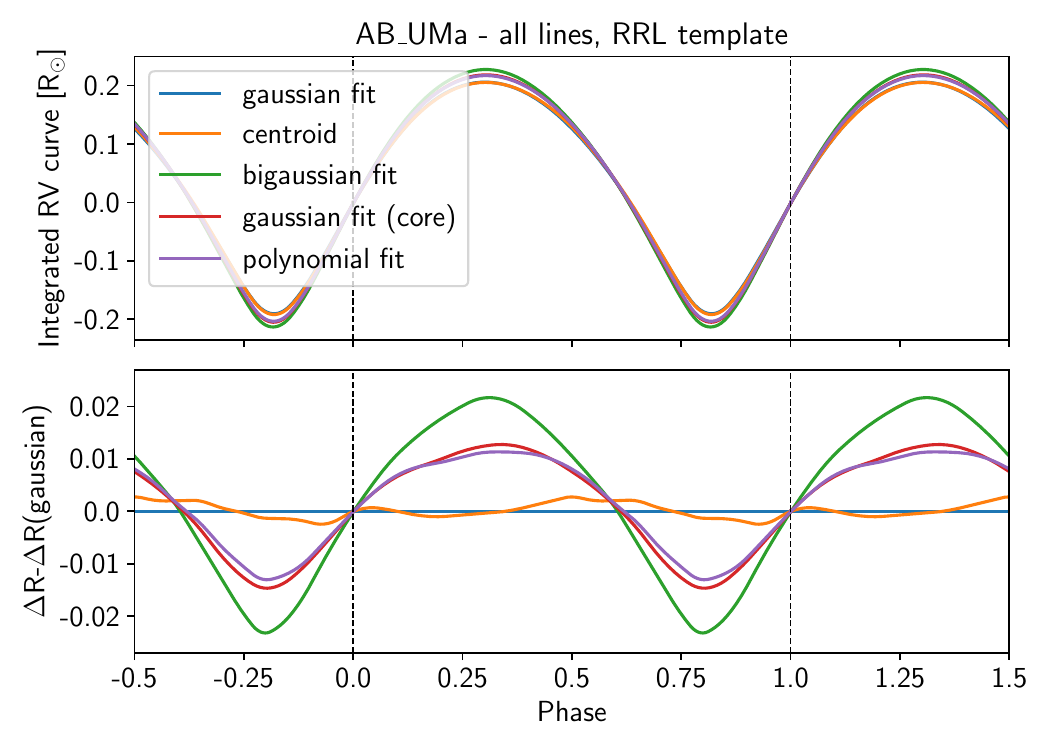}
   \caption{Comparison of the impact of different CCF measurement methods on the radius curve of AB UMa (the RV curve is shown in Fig.~\ref{def_comparison_RV}). The bottom plot shows the difference from the more commonly used method, that is, the fit of a Gaussian profile to the CCF.}
   \label{def_comparison_intRV}
\end{figure}

 Here we present the effect of the choice of method on the RV curve (plotted in Fig.~\ref{def_comparison_RV}), and the resulting difference in the amplitude of the integrated RV curve (plotted in Fig.~\ref{def_comparison_intRV}). Comparisons of RV curves using different techniques have been carried out for single lines \citep[e.g.][]{Nardetto2006}, and on CCF for Cepheids \citep{Anderson2016} with the conclusion that there is a non-negligible difference in the RV curves when using one or the other method (i.e. Gaussian, centroid, bi-Gaussian etc.).
We studied five definitions: the mean of a Gaussian fit, the mean of a bi-Gaussian fit, the mean of a Gaussian fit of the core of the CCF (defined as containing approximately 68$\%$ of the CCF absorption), the centroid, and the minimum using a fourth-order polynomial fit. For AB UMa, the difference in the amplitude of the curve ranges from $0.2\%$ (centroid method) to $11\%$ (bi-Gaussian fit) of the amplitude compared to the classic Gaussian fit method. Based on these differences, we estimate that the choice of the definition of the RV given a CCF can lead to a change in the $p$-factor of up to 10\%.

As a summary of this analysis, Table~\ref{tab:error_p} provides the major sources of variation in the amplitude of the integrated RV curve for each aspect studied, and for two different stars.
We conclude that the main source of variation of the $p$-factor is caused by the choice of the CCF measurement method ---rather than the choice of mask or template, the code used to compute the CCF, or even the line selection.
It therefore appears that the choice of CCF measurement method can significantly affect the RV, and therefore also the $p$-factor.

For the remainder of our analysis and the determination of the $p$-factor, we used the mean of the Gaussian fit of the CCF, as this is the most common method but is also the most precise with a poor S/N \citep{Anderson2016}. This allows us to directly compare our values of the $p$-factor with previous determinations from the literature.

\begin{figure}
   \centering
   \includegraphics[width=\hsize]{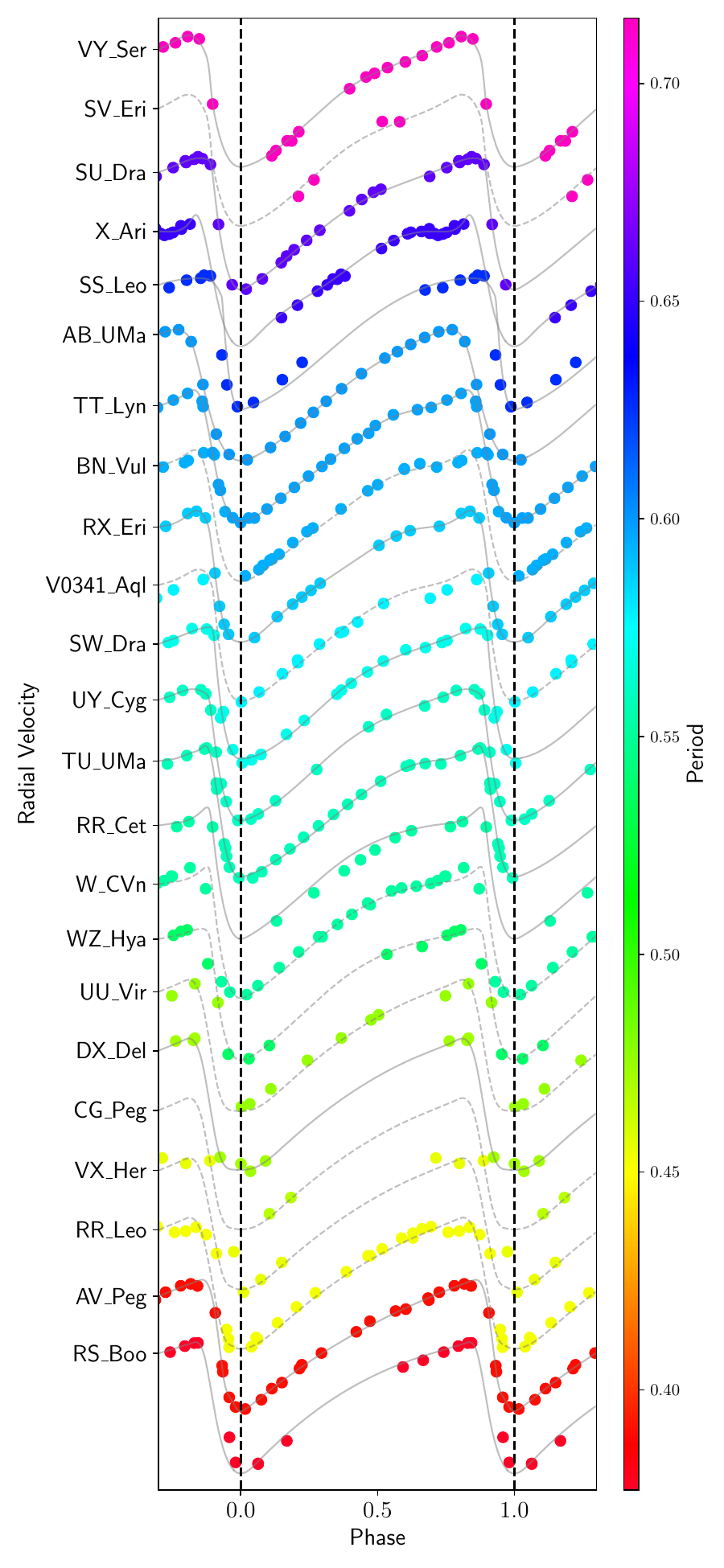}
   \caption{All RV curves obtained with the SOPHIE spectrograph. The curves have been normalised using \textit{Gaia} parameters if an insufficient number of measurements was obtained. An offset in velocity is added to see all curves in a single plot, which is related to the period.}
   \label{all_RV_SOPHIE}
\end{figure}

All RV curves are plotted in Fig.~\ref{all_RV_SOPHIE}, and are colour-coded according to pulsation period. SPIPS-modelled curves are shown in grey (see details in Sect.~\ref{SPIPS}). For stars without a model, a dashed curve is plotted, which is the weighted mean of models of the most similar stars in terms of period (of those for which a model has been determined). This is interesting as it reveals the continuity of the shape of RV curves with changing period. 

\begin{table}
\caption{Summary of the impact on the amplitude of the integrated RV curve when using different methods to determine RV for AB UMa and TT Lyn.}             %
\label{tab:error_p}      %
\centering                          %
\begin{tabular}{c c c}        %
\hline\hline                 %
Source of error & AB UMa & TT Lyn \\    %
\hline                        %
   Code CCF & 0.81\% & 0.67\% \\      %
   Masks/Templates & 1.40\% & 1.11\% \\
   Line selection & 1.42\% & 0.35\% \\
   CCF measurement method & 11.17\% & 8.35\% \\
\hline                                   %
\end{tabular}
\end{table}

\begin{table*}
\caption{Measurements of RV from spectra with the SOPHIE spectrograph. We provide values from a Gaussian (RV$_{g}$) and a bi-Gaussian (RV$_{2g}$) fit of the CCF of spectra with the computed binary mask containing all unblended lines and from a Gaussian fit of the CCF of spectra with the computed binary masks containing deep lines (RV$_{\mathrm{deep}}$) or medium lines (RV$_{\mathrm{medium}}$). The errors come from the fit of the Gaussian or bi-Gaussian fit and may be underestimated. This table is available in its entirety in the CDS.}
\label{tab_RV}
\centering
\begin{tabular}{lrrrrrrrrrr}
\hline
\hline
 Star  &  HJD & S/N & RV$_{g} $ & $\sigma_{\mathrm{RV}_g}$ & RV$_{2g}$ & $\sigma_{\mathrm{RV}_{2g}}$ & RV$_{\mathrm{deep}}$ & $\sigma_{\mathrm{RV}_{\mathrm{deep}}}$ & RV$_{\mathrm{medium}}$ & $\sigma_{\mathrm{RV}_{\mathrm{medium}}}$ \\
\hline
 V0341 Aql & 2459756.5572 & 25.2 & -114.86 & 0.11 & -119.67 & 0.52 & -115.16 & 0.05 & - & - \\
 V0341 Aql & 2459756.5951 & 15.2 & -109.72 & 0.12 & -110.11 & 0.56 & -109.50 & 0.06 & -108.99 & 0.17 \\
 V0341 Aql & 2459756.4765 & 10.6 & -51.49 & 0.03 & -49.53 & 0.17 & -50.92 & 0.02 & -51.12 & 0.07 \\
\hline
\end{tabular}
\end{table*}

\section{The SPIPS modelling}\label{SPIPS}
\subsection{The SPIPS code}

The Spectro-Photo-Interferometry for Pulsating Stars \citep[SPIPS,][]{SPIPS} is a modelling code that simultaneously fits a broad range of observables of a pulsating star to determine its physical parameters (pulsation period, bolometric luminosity, distance or projection-factor, effective temperature, radius, reddening, etc.).
SPIPS allows us to simultaneously fit different types of observations for a single star: RVs, spectroscopic effective temperatures, angular diameters from interferometry and photometric magnitudes in all bands and colours.
Firstly, an interpolated curve is fitted to the RV and another for temperature data. Then, the other observables are predicted as a function of pulsation using the fitted parameters, and the photometry is modelled thanks to a grid of static ATLAS9 models \citep{ATLAS9} for the given temperature, metallicity, and surface gravity. The latter is estimated from the star's radius and the period--mass--radius relation from \cite{Bono2001}. Such relations are calibrated for Cepheids but surface gravity does not have a large influence on the large-band photometry, with an impact on the $p$-factor of $\lesssim 1\%$. Filter transmissions and photometric zero points are used to compare with observations in a given set of filters.

\subsection{The modelling strategy}

To determine the $p$-factor  for our RRLs, we adopted the distances from \citet{BailerJonesCatalog} adapted from the \textit{Gaia} DR3 catalogue \citep{GaiaMission,GaiaDR3} and considering the zero-point correction on parallax given by \cite{Lindegren2021}. The distances of the stars in our sample range between 536 and 1258 pc. For such nearby stars, errors on the parallax are quite small, and therefore one could also choose the distance as the inverse of the parallax. We checked that this choice does not affect the result. The impact is around 2\% on the $p$-factor, which is quite similar to the uncertainty on this value. As we do not have spectroscopic temperature measurements for all stars, we adopted the colour excess $E(B-V)$ predicted by the Stilism 3D maps using the same distances \citep{Stilism}. Metallicities were set to the homogeneous values reported in \cite{MonsonPhotRRL} \citep[which include original values from][]{FernleyMetal1998, FernleyBarnesMetal1997, fernleyMetalBBPup1998}, and we considered a typical uncertainty of $\pm$0.15 dex.
Because of the large number of parameters in the global fit, we individually checked the convergence of the optimisation to convincing solutions. As stars in our sample are relatively close and bright, they have been well observed and realistic initial parameters are easy to find.
We chose to initialise the period, the gamma velocity, and the amplitude of the RV curve to the \textit{Gaia} values found by  \cite{Clementini2023GAIARRL}. The starting value of the star's radius is defined according to the initial period thanks to the period--radius relation of \cite{Marconi2015}.

It is important to note that the various data used are separated by up to several decades, as many photometric measurements were collected in the 1980s. Such an interval corresponds to tens of thousands of pulsation cycles. Therefore, the fit is highly sensitive to the period. This is why we only fit the period in the last step, allowing only very small variations.
We also set the reference time to be in the average of all the data in order to avoid adding the errors as many times as the number of cycles between the older and the more recent observation. Finally, we occasionally considered the temporal change in the pulsation period as an additional parameter, depending on the goodness of the fit.

Here is the strategy we adopted to model the pulsation cycle of a typical star in our sample. First, we only fit the effective temperature curve with temperature measurements (if any) and photometry. The curve is fitted using Fourier series with as many harmonics as necessary to properly reproduce the data (between 6 and 8). Photometric colours are not used for the general fit as we favoured single-band photometric measurements.
In a second step, we only fit the RV model curve to RV measurements. Because the phase coverage is generally not very dense, Fourier series are not suitable for RV curves as they introduce small oscillations. We therefore used cubic splines, with between six and eight nodes.
The choice of the number of Fourier series or nodes of splines to use was made through a visual inspection of the model and of the individual and global $\chi^2$.
Given these first guesses of both RV and temperature curves, we can then fit all the parameters but the period, before a finer and final tuning including the period determination. Metallicities, colour excesses, and distances are fixed during the whole procedure, but the impact of each of these parameters has been considered in the final $p$-factors and the associated uncertainties.

Because we were not able to obtain a good model with this procedure  for some stars, we then adapted the code individually for each star to reach a good result in the fitting curves, that is, for RS Boo, AV Peg, and TU UMa. 
For these three stars, because of an observed phase shift between the model and some datasets that was not reduced by simply adding a period change, we performed a first model based on contemporaneous data and used the fitted parameters as initial parameters for a second run, including the other datasets. This helped the fitting process to converge. For RS Boo, the initial parameters were computed on data exclusively from \cite{JonesBW1988}. For AV Peg we used the \cite{MonsonPhotRRL} dataset together with the OHP/SOPHIE RV curve and \cite{AndTeff1} temperature for the first step. For TU UMa, the initial parameters were obtained using \cite{MonsonPhotRRL} and \textit{Gaia} photometry, the OHP/SOPHIE RV curve, and the temperature provided by \cite{AndTeff1} . In the particular case of RS Boo, its Blazhko effect prevented us from fitting one observable with various datasets, and so the RV curve from OHP/SOPHIE is not fitted. This leads to a higher uncertainty in the fit, especially for the period determination.

We found a good model for eight RRLs with new measurements taken at OHP using a Gaussian fit of the cross-correlation with a binary mask that includes all the identified unblended lines in a synthetic spectrum.
We completed this sample with six additional stars from our list, using literature RVs that were offset to match our SOPHIE RV measurements (such an additive offset does not affect the $p$-factor). We also added three more southern stars using exclusively data from the literature.

\begin{figure*}
        \centering
        \includegraphics[width=\hsize]{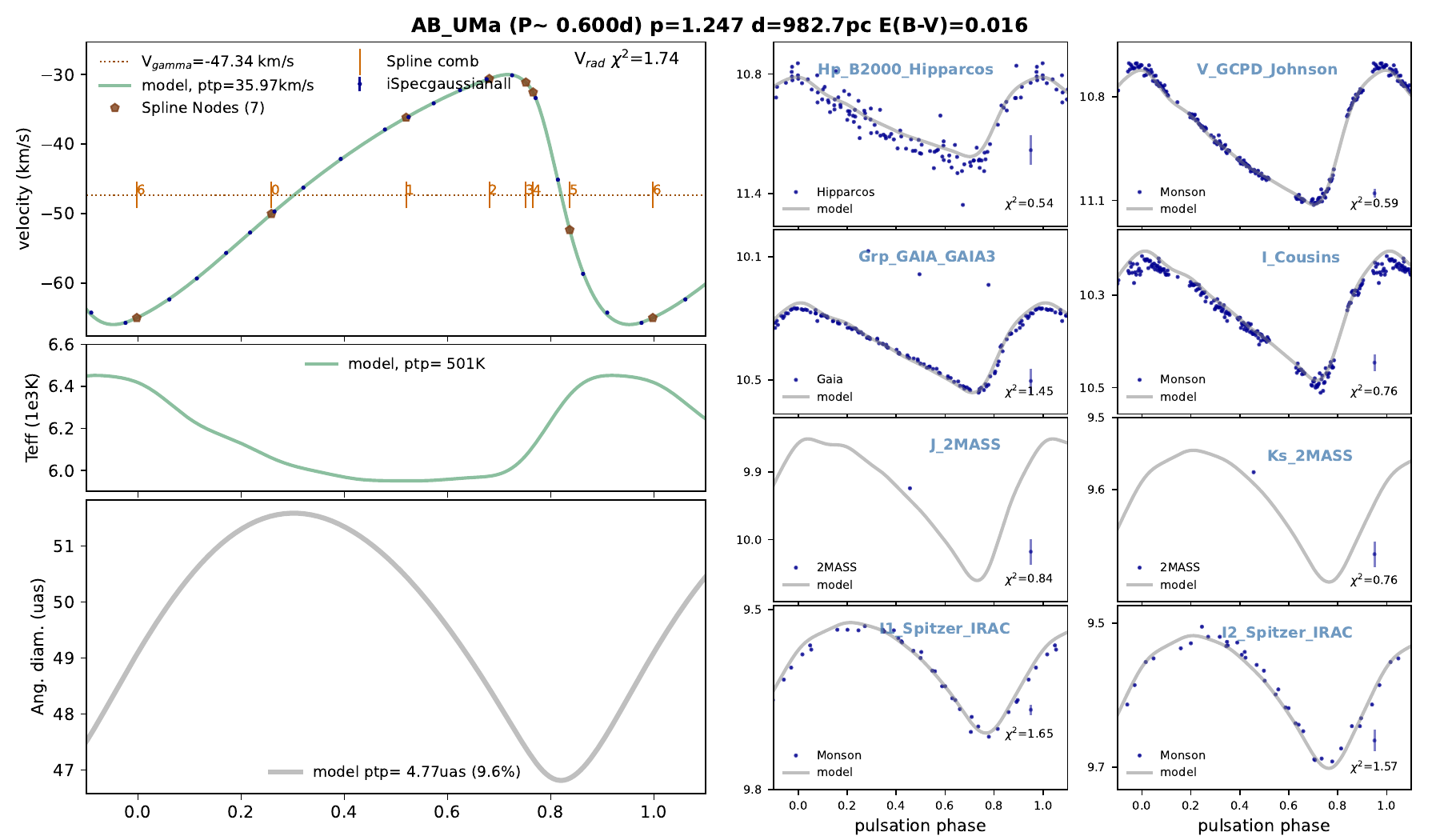}
        \caption{SPIPS model of AB UMa with RV from SOPHIE spectra (Gaussian fit of the CCF using the binary mask built for RRL with all unblended lines). Photometry measurements are from \cite{MonsonPhotRRL} (60\% of available photometric measurements).}
        \label{SPIPS_AB_UMa}
\end{figure*}

\begin{figure*}
        \centering
        \includegraphics[width=\hsize]{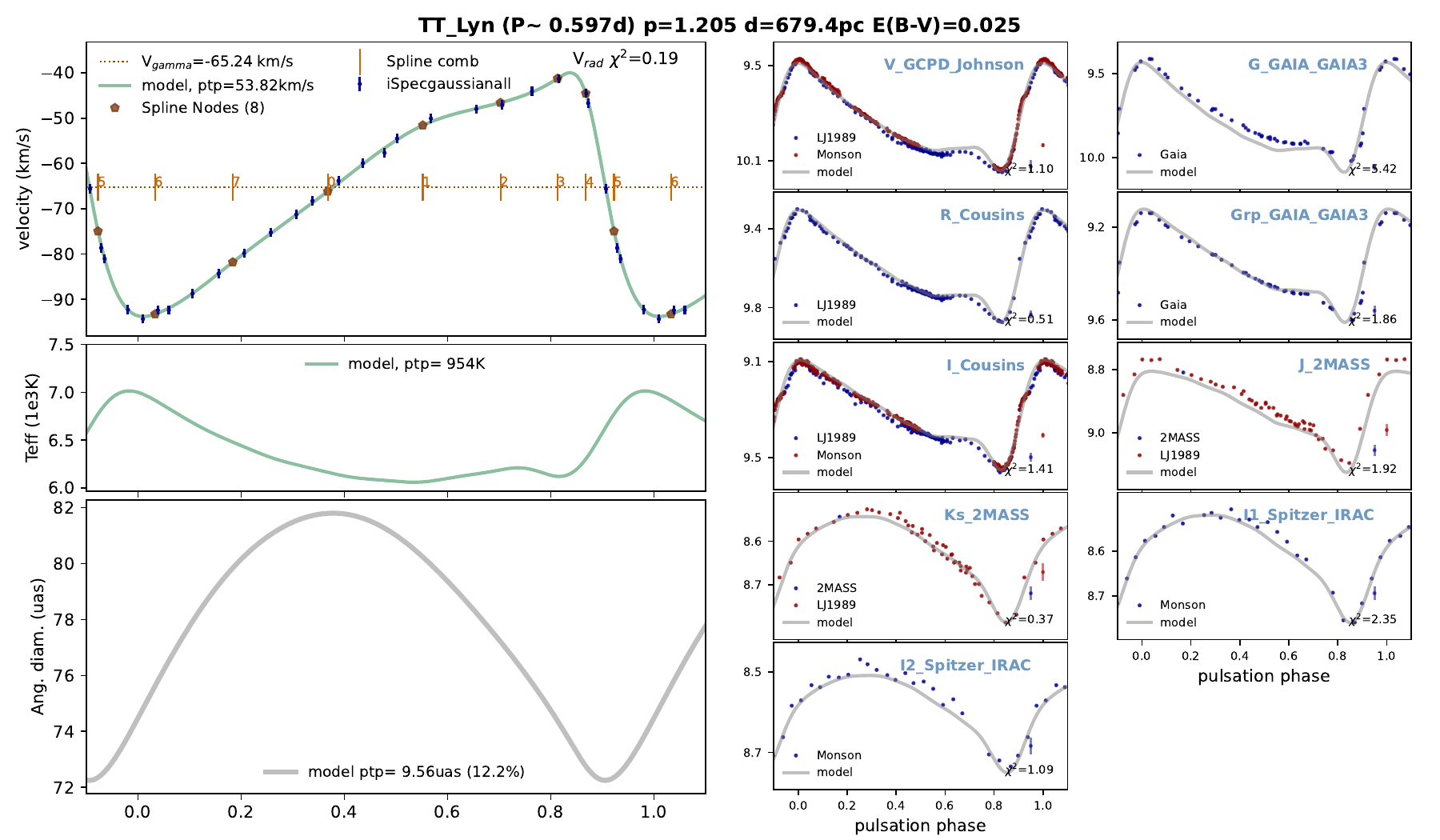}
        \caption{SPIPS model of TT Lyn with RV from SOPHIE spectra (Gaussian fit of the CCF using the binary mask built for RRLs with all unblended lines). Photometry measurements are from \cite{MonsonPhotRRL} and \cite{LJBW1989} (29\% of available photometric measurements).}
        \label{SPIPS_TT_Lyn}
\end{figure*}

Figures~\ref{SPIPS_AB_UMa} and \ref{SPIPS_TT_Lyn} present the result of the SPIPS modelling of two RRLs with new SOPHIE RV measurements. Others are provided in Appendix~\ref{appendix:SPIPS}\footnote{Models are all available at the CDS}.
The plot shows velocity, temperature, angular diameter, and photometric curves, both observed (points) and modelled (solid lines). The model curves shown in grey are computed based on the green curves whose parameters are optimised during the fitting process to fit the various observables simultaneously. For the photometric curves, the filter is indicated in blue.

\section{Results}\label{results}

In total, 17 RRLs have been modelled using the SPIPS code. The properties of the modelled RRLs are summarised in Table~\ref{tab:SPIPS_properties}.

\renewcommand{\arraystretch}{2}
\begin{table*}
\caption{Main properties of the best-fit models for 17 field RRLs.
The distance ($d$), metallicity ([Fe/H]), and colour excess $E(B-V)$ are fixed. We fit the period ($P$ with uncertainty $\sigma_P$), if needed a change in period ($\dot P$), and the $p$-factor. From the RV and photometry curves, we determine the mean radius $R$, the mean effective temperature $T_\mathrm{eff}$, and the gamma velocity $V_\gamma$. The three groups correspond to stars with SOPHIE RV data exclusively, stars with some additional literature RV data, and stars with literature RV data only, respectively.}             
\label{tab:SPIPS_properties}      
\centering         
\resizebox{\textwidth}{!}{%
\begin{tabular}{c c c c c c c c c c c c}     
\hline\hline
Star & $d$ (pc) & [Fe/H] & $E(B-V)$ & $P$ (h) & $\sigma_P$ (h) & $\dot P$ (ms.yr$^{-1}$) & $R$ ($R_\odot$) & $T_\mathrm{eff}$ (K) &  $V_\gamma$ ($\mathrm{km.s}^{-1}$)  & $p$ & $\chi^2_r$\\
\hline
UY Cyg & 1039$_{-23}^{+22}$ & -0.80 & 0.149$_{-0.026}^{+0.026}$ & 13.4569641 & 28$.10^{-7}$ & - & 5.14$_{\pm0.09}$ & 6658$_{\pm8}$ & 7.62$_{\pm3.01}$ & 1.17$_{\pm0.04}$ & 2.8 \\ 
SU Dra & 741$_{-7}^{+9}$ & -1.80 & 0.022$_{-0.014}^{+0.000}$ & 15.8501639 & 12$.10^{-7}$ & - & 6.15$_{\pm0.15}$ & 6458$_{\pm4}$ & -164.74$_{\pm0.95}$ & 1.20$_{\pm0.03}$ & 2.8 \\ 
SW Dra & 930$_{-16}^{+17}$ & -1.12 & 0.022$_{-0.010}^{+0.000}$ & 13.6721226 & 8$.10^{-7}$ & 0.88$_{\pm0.64}$ & 5.58$_{\pm0.11}$ & 6421$_{\pm8}$ & -29.05$_{\pm0.98}$ & 1.22$_{\pm0.04}$ & 2.0 \\ 
RX Eri & 580$_{-7}^{+7}$ & -1.33 & 0.031$_{-0.013}^{+0.013}$ & 14.0939041 & 18$.10^{-7}$ & - & 5.54$_{\pm0.18}$ & 6214$_{\pm5}$ & 67.90$_{\pm1.03}$ & 1.25$_{\pm0.02}$ & 1.8 \\ 
TT Lyn & 679$_{-8}^{+7}$ & -1.56 & 0.025$_{-0.013}^{+0.000}$ & 14.3383528 & 15$.10^{-7}$ & -3.76$_{\pm0.77}$ & 5.70$_{\pm0.16}$ & 6359$_{\pm8}$ & -65.24$_{\pm0.86}$ & 1.20$_{\pm0.03}$ & 1.6 \\ 
AV Peg & 665$_{-10}^{+5}$ & -0.08 & 0.062$_{-0.003}^{+0.003}$ & 9.3691917 & 6$.10^{-7}$ & 9.36$_{\pm0.51}$ & 3.87$_{\pm0.11}$ & 6603$_{\pm6}$ & -56.07$_{\pm0.53}$ & 1.32$_{\pm0.03}$ & 0.9 \\ 
AB UMa & 983$_{-17}^{+18}$ & -0.49 & 0.016$_{-0.015}^{+0.011}$ & 14.3899565 & 28$.10^{-7}$ & - & 5.25$_{\pm0.10}$ & 6137$_{\pm5}$ & -47.34$_{\pm0.10}$ & 1.25$_{\pm0.04}$ & 1.1 \\ 
TU UMa & 626$_{-11}^{+11}$ & -1.51 & 0.018$_{-0.022}^{+0.002}$ & 13.3837597 & 5$.10^{-7}$ & 1.71$_{\pm0.02}$ & 5.16$_{\pm0.15}$ & 6407$_{\pm6}$ & 96.69$_{\pm0.69}$ & 1.21$_{\pm0.06}$ & 1.9 \\ 
\hline
X Ari & 536$_{-6}^{+5}$ & -2.43 & 0.254$_{-0.043}^{+0.003}$ & 15.6279509 & 21$.10^{-7}$ & 16.39$_{\pm0.78}$ & 6.20$_{\pm0.21}$ & 6789$_{\pm16}$ & -36.00$_{\pm1.01}$ & 1.29$_{\pm0.04}$ & 1.5 \\ 
RS Boo & 745$_{-10}^{+9}$ & -0.36 & 0.019$_{-0.019}^{+0.000}$ & 9.0561197 & 32$.10^{-7}$ & 47.36$_{\pm0.20}$ & 4.14$_{\pm0.10}$ & 6700$_{\pm36}$ & -3.61$_{\pm0.70}$ & 1.19$_{\pm0.07}$ & 4.8 \\ 
RR Cet & 610$_{-8}^{+10}$ & -1.45 & 0.028$_{-0.022}^{+0.003}$ & 13.2726998 & 8$.10^{-7}$ & - & 5.30$_{\pm0.16}$ & 6408$_{\pm5}$ & -71.80$_{\pm1.86}$ & 1.32$_{\pm0.03}$ & 1.6 \\ 
DX Del & 568$_{-5}^{+5}$ & -0.39 & 0.067$_{-0.024}^{+0.024}$ & 11.3428460 & 3$.10^{-7}$ & - & 4.59$_{\pm0.15}$ & 6429$_{\pm4}$ & -56.43$_{\pm0.72}$ & 1.30$_{\pm0.03}$ & 1.8 \\ 
SS Leo & 1258$_{-36}^{+40}$ & -1.79 & 0.023$_{-0.023}^{+0.005}$ & 15.0321238 & 17$.10^{-7}$ & - & 5.74$_{\pm0.08}$ & 6498$_{\pm12}$ & 163.40$_{\pm0.60}$ & 1.06$_{\pm0.05}$ & 2.5 \\ 
VY Ser & 827$_{-15}^{+16}$ & -1.79 & 0.045$_{-0.037}^{+0.005}$ & 17.1383252 & 38$.10^{-7}$ & -0.38$_{\pm0.19}$ & 6.55$_{\pm0.15}$ & 6241$_{\pm11}$ & -144.48$_{\pm0.81}$ & 1.39$_{\pm0.06}$ & 1.9 \\ 
\hline
WY Ant & 1019$_{-20}^{+22}$ & -1.48 & 0.042$_{-0.012}^{+0.012}$ & 13.7842117 & 6$.10^{-7}$ & - & 5.43$_{\pm0.10}$ & 6370$_{\pm5}$ & 203.53$_{\pm0.27}$ & 1.12$_{\pm0.03}$ & 1.9 \\ 
RV Oct & 978$_{-11}^{+12}$ & -1.71 & 0.105$_{-0.019}^{+0.019}$ & 13.7080606 & 4$.10^{-7}$ & - & 5.57$_{\pm0.11}$ & 6306$_{\pm6}$ & 141.01$_{\pm0.20}$ & 1.46$_{\pm0.04}$ & 1.9 \\ 
BB Pup & 1604$_{-39}^{+43}$ & -0.64 & 0.063$_{-0.064}^{+0.024}$ & 11.5331719 & 5$.10^{-7}$ & - & 4.70$_{\pm0.05}$ & 6390$_{\pm7}$ & 132.65$_{\pm0.90}$ & 1.35$_{\pm0.09}$ & 1.5 \\ 
\hline
\end{tabular}}
\end{table*}

\subsection{$p$-factor}

We compare our determinations of p-factors (listed in Table~\ref{tab:SPIPS_properties}) with those of several previous Baade-Wesselink analysis ---in terms of the ratio between
the distance and the $p$-factor $d/p$.
All these studies adopted a constant $p$-factor of between 1.3 and 1.36 to determine the distances of RRLs and their absolute magnitudes. A detailed comparison is provided in Table~\ref{tab:d/p}, and we summarise our main conclusions below.

\begin{itemize}
\item We note that our results for the ratio $d/p$ are in good agreement with those of \citet{FernleyBWXAri1989}, \citet{SkillenDXDel1989}, \citet{FernleySSLeoVYSer1990}, \citet{SkillenWYAntRVOctBBPup1993} for X Ari, DX Del, SS Leo, VY Ser, and RV Oct, but not for BB Pup and WY Ant, and are also in agreement with the results of \citet{LJAbsMag1990} for RR Cet, RX Eri, AV Peg, and TU UMa, but not for TT Lyn and SW Dra.
\item For the stars present in the studies by \cite{JonesBWXAri1987,JonesBWSWDra1987,JonesBW1988} (X Ari, SW Dra, VY Ser), our results do not agree in terms of the $d/p$ ratio but the results for X Ari and VY Ser in the mentioned papers are also not consistent with the ones in \citet{FernleyBWXAri1989,FernleySSLeoVYSer1990}, closer to the results of the present paper.
\item Similarly, our results for RR Cet and DX Del are comparable with those in \citet{SkillenDXDel1989} and \citet{LJAbsMag1990}, but differ from the results of  \citet{BurkiRRCet1986,BurkiDXDel1986}. 
\item For SW Dra, we find a different ratio $d/p$ from all other authors \citep{JonesBWSWDra1987,CacciariBWSWAndSWDraSSFor1989}.
\item Finally, it is worth noting that, even if they agree within the error bars, ratios $d/p$ derived from the SPIPS algorithm are almost systematically higher than those in the literature, leading to smaller $p$-factors.
\end{itemize}

\renewcommand{\arraystretch}{1}

\begin{table*}
\caption{Comparison of results for the ratio $d/p$ from the present study (second column) and previous determinations (last column). Stars indicate the fixed parameter for determinations from other studies.}
\label{tab:d/p}
\centering
\begin{tabular}{c c c c c c}
\hline\hline
Name & $d_{Bailer-Jones}/p_{\mathrm{SPIPS}}$ & Reference BW & $d_{BW}$ (pc) & $p_{BW}$ & $d_{BW}/p_{BW}$ \\
\hline
WY Ant & 910$\pm$31 & \cite{SkillenBW1993} & 1034$\pm$10 & 1.33$^*$ & 777$\pm$8\\
\hline
 & & \cite{FernleyBWXAri1989} & 505$\pm$30 & 1.33$^*$ & 380$\pm$23\\
X Ari & 416$\pm$14 & \cite{JonesBWXAri1987} & 440$\pm$30 & 1.30$^*$ & 338$\pm$23 \\
 &  & \cite{JonesBW1988} & 466$\pm$13 & 1.30$^*$ & 358$\pm$10\\
\hline
RS Boo & 626$\pm$38 & \cite{JonesBW1988} & 741$\pm$22 & 1.30$^*$ & 570$\pm$17\\
\hline
RR Cet & 462$\pm$13 & \cite{BurkiRRCet1986} & 760$\pm$40 & 1.36$^*$ & 559$\pm$29\\
 &  & \cite{LJAbsMag1990} & 587$\pm$32 & 1.32$^*$ & 445$\pm$24\\
\hline
DX Del & 437$\pm$11 & \cite{BurkiDXDel1986} & 750$\pm$45 & 1.36$^*$ & 559$\pm$29\\
 & & \cite{SkillenDXDel1989} & 593$\pm$50 & 1.33$^*$ & 446$\pm$38\\
 \hline
SU Dra & 618$\pm$17 & \cite{LJAbsMag1990} & 640$\pm$35 & 1.32$^*$ & 485$\pm$27\\
\hline
 &  & \cite{CacciariBWSWAndSWDraSSFor1989} & 840$\pm$50 & 1.36$^*$ & 618$\pm$37\\
SW Dra & 762$\pm$29 & \cite{JonesBWSWDra1987} & 824$\pm$40 & 1.30$^*$ & 634$\pm$31\\
 &  & \cite{JonesBW1988} & 836$\pm$23 & 1.30$^*$ & 643$\pm$18\\
 \hline
RX Eri & 464$\pm$9 & \cite{LJAbsMag1990} & 568$\pm$31 & 1.32$^*$ & 430$\pm$23\\
\hline
SS Leo & 1187$\pm$68 & \cite{FernleySSLeoVYSer1990} & 1414$\pm$80 & 1.33$^*$ & 1063$\pm$60\\
\hline
TT Lyn & 566$\pm$16 & \cite{LJAbsMag1990} & 654$\pm$36 & 1.32$^*$ & 495$\pm$27\\
\hline
RV Oct & 670$\pm$20 & \cite{SkillenBW1993} & 904$\pm$20 & 1.33$^*$ & 680$\pm$15\\
\hline
AV Peg & 504$\pm$14 & \cite{LJAbsMag1990} & 651$\pm$36 & 1.32$^*$ & 493$\pm$27\\
\hline
BB Pup & 1188$\pm$85 & \cite{SkillenBW1993} & 1328$\pm$20 & 1.33$^*$ & 998$\pm$15\\
\hline
VY Ser & 595$\pm$28 & \cite{FernleySSLeoVYSer1990} & 750$\pm$90 & 1.33$^*$ & 564$\pm$68\\
 &  & \cite{JonesBW1988} & 682$\pm$19 & 1.30$^*$ & 525$\pm$15\\
 \hline
TU UMa & 517$\pm$27 & \cite{LJAbsMag1990} & 621$\pm$34 & 1.32$^*$ & 470$\pm$26\\

   \hline
\end{tabular}
\end{table*}

\begin{figure}
   \centering
   \includegraphics[width=\hsize]{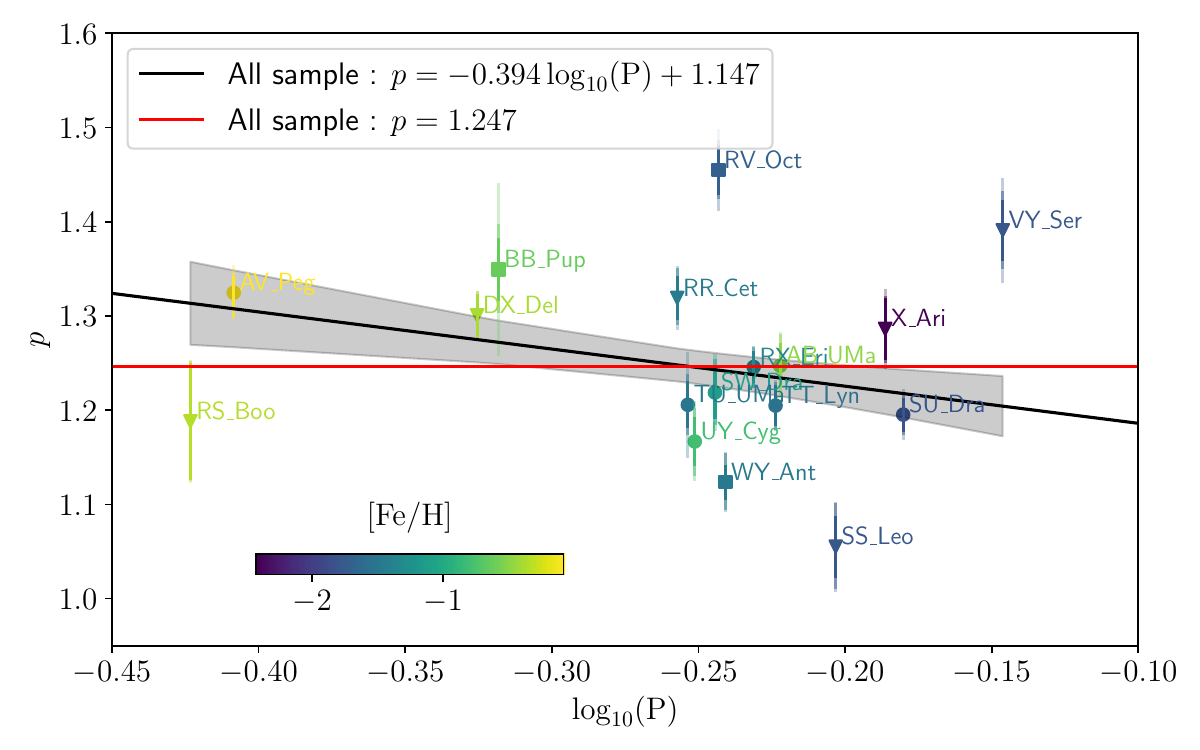}
   \caption{$p$-factors for RRLs from the SPIPS model, fitted with a linear function in black and a constant in red. Points are for stars with an RV curve from the OHP exclusively. Inverted triangles represent stars for which the RV curve from the literature has been rescaled to OHP measurements. Finally, squares are the three additionnal stars for which all data are from the literature.}
   \label{pP_RRL}
\end{figure}

The resulting $p$-factors of the 17 RRLs (defined for a Gaussian fit of the CCF) are plotted against the pulsation period in Fig.~\ref{pP_RRL}, encoded according to the origin of the RV measurements, and together with a linear fit and a constant fit. We find a mean value of $p= 1.248 \pm 0.022$

\begin{figure}
   \centering
   \includegraphics[width=\hsize]{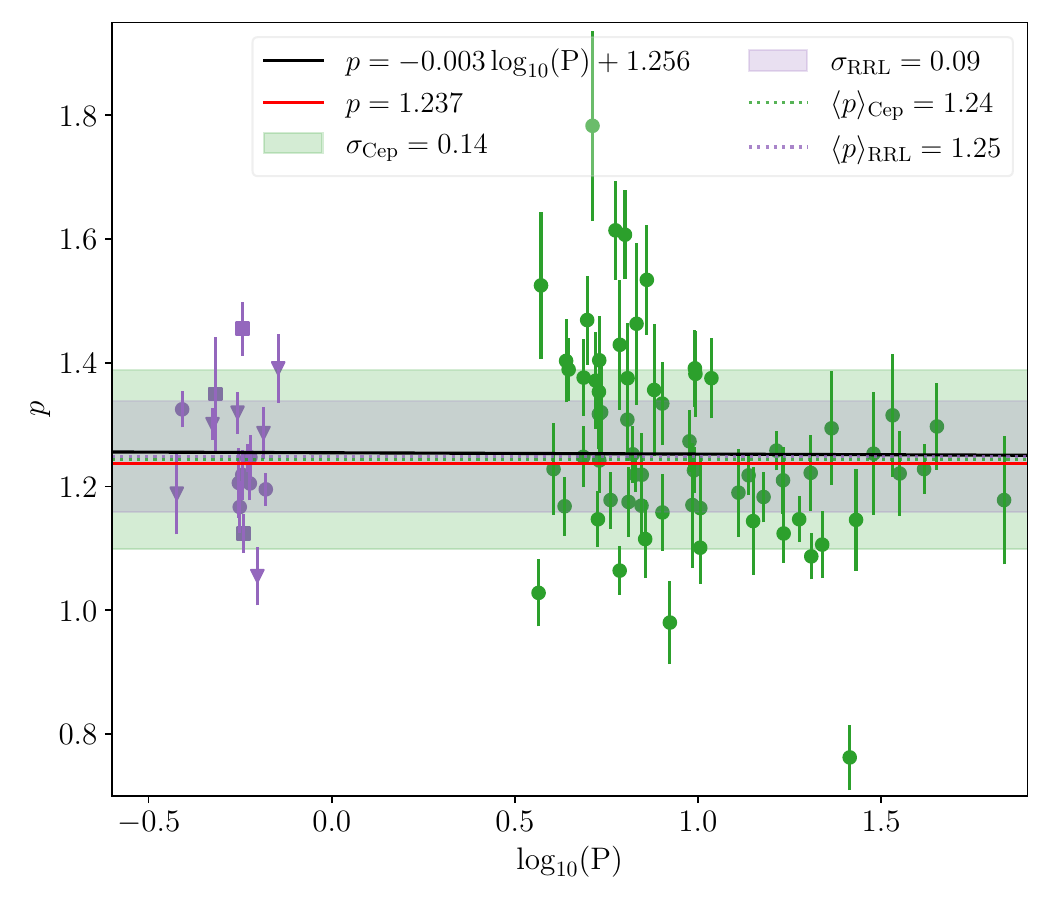}
   \caption{Relation between the $p$-factor and the pulsation period for RRLs in this study and Cepheids from \cite{TrahinpCep}, who used the same tool for the modelling.}
   \label{pP_RRLCep}
\end{figure}

In Fig.~\ref{pP_RRLCep}, we compare the dispersion of the RRL $p$-factors with that of the Cepheids studied by \cite{TrahinpCep}, who also used \textit{Gaia} EDR3 parallaxes. The scatter for RRLs is smaller (7\%) than for Cepheids (12\%).
It would be interesting to compare this to other classes of pulsating stars like $\delta$ Scuti (\cite{Nardetto2014} found higher $p$-factors with quite low dispersion of 2\% for a sample of four stars) or type II Cepheids for instance (work by P.Wielgórski, in preparation).
However, the scatter in $p$ for RRLs remains too high to use the PoP technique to measure their distances with an accuracy of 1\%.

Our $p$-factor distribution is compatible with a constant $p$-factor for all stars (RRLs and Cepheids) of $p=1.245 \pm 0.015$, which is uncorrelated with period. This constant value is compatible with previous results obtained with the SPIPS algorithm for Cepheids in the Magellanic Clouds and in the Milky Way \citep{Gallenne2017}, and is relatively close to the result of \cite{Breitfelder2016}.

\begin{figure*}
        \centering
        \includegraphics[width=\hsize]{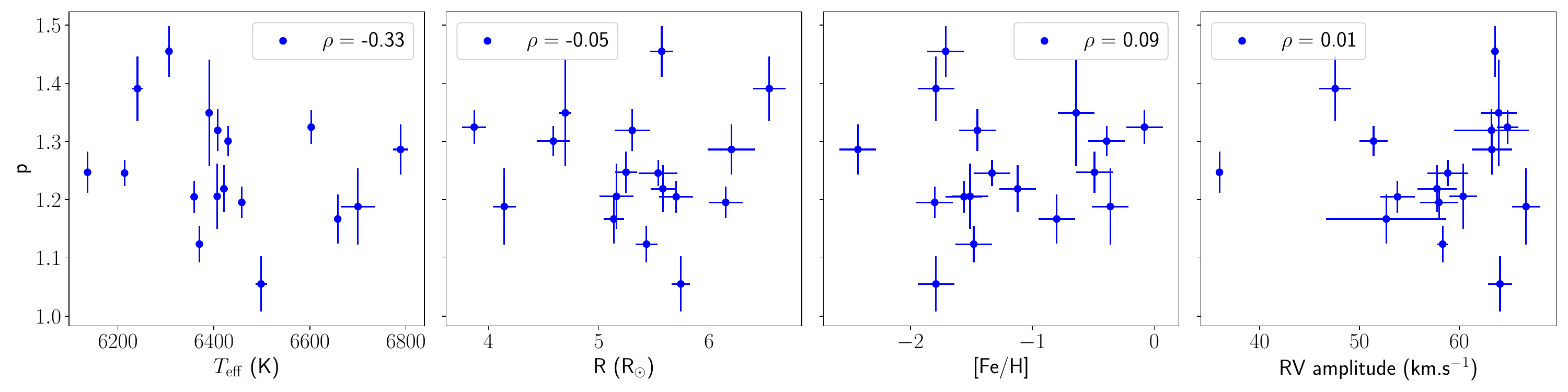}
        \caption{Comparison between the $p$-factor of 17 RRLs and effective temperature, mean radius, metallicity, and amplitude of the RV curve. For each plot, the Spearman's coefficient is given to test the statistical probability of a monotonic relation between the $p$-factor and the other parameters.}
        \label{p_params}
\end{figure*}

Comparison of the $p$-factor of the RRLs in our sample with some of their parameters (see Fig.~\ref{p_params}) did not reveal any correlation, that is, with the effective temperature, the radius, the metallicity, or the amplitude of the RV curve. A more complex dependence is not excluded but further analysis would be necessary to investigate this properly.

\subsection{Period--luminosity relation}

\begin{figure}
   \centering
   \includegraphics[width=\hsize]{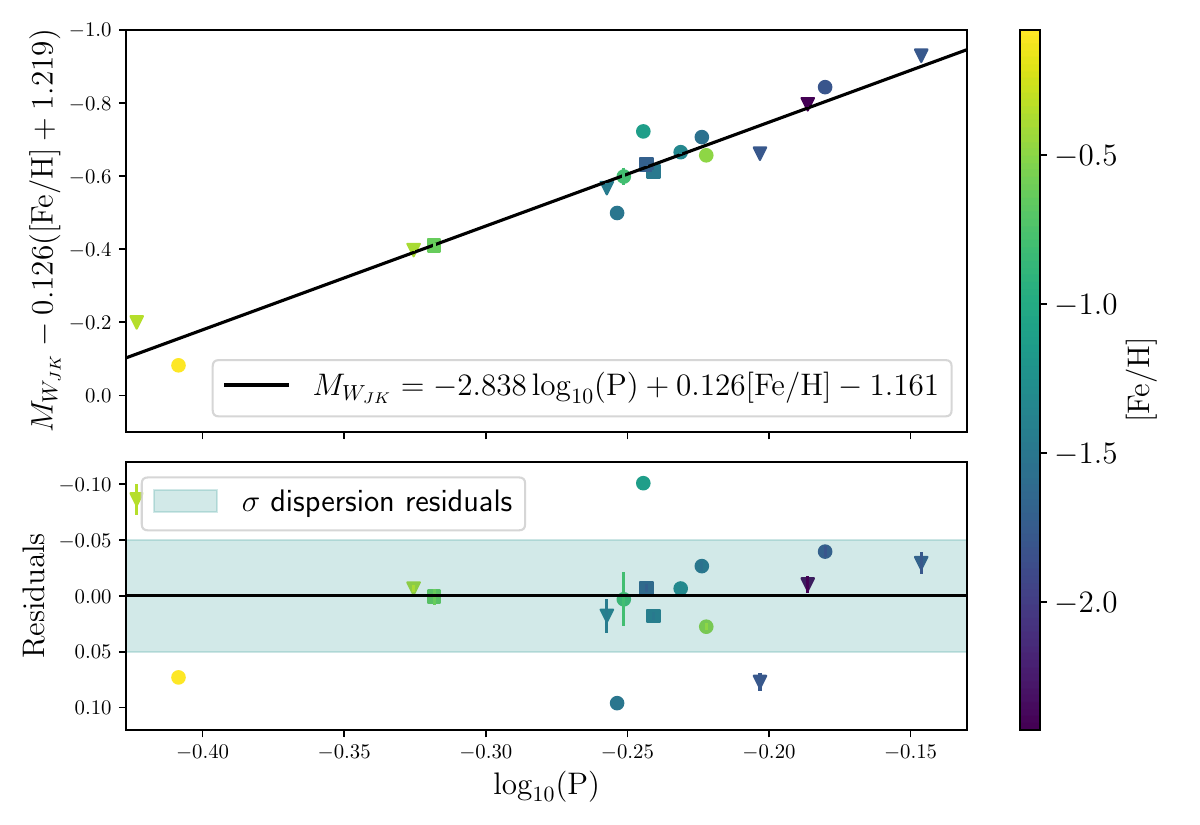}
   \caption{Period--luminosity--metallicity relation in the Wesenheit index for the 17 RRLs of our sample. A plane has been fitted to the data and we show here the magnitude corrected for the metallicity effect versus the logarithm of the period.}
   \label{PLZR}
\end{figure}

From the SPIPS models, we also have access to the main parameters of the star, which allows us to determine period--radius and period--luminosity relations. While period--luminosity relations are the principal tool for distance determinations, they are not always defined for RRLs \citep{Bhardwaj2023}. Figure~\ref{PLZR} shows this latter relation for the dereddened absolute magnitude in the Wesenheit index $W_{JK} = K_S-0.756(J-K_S)$ (using the reddening law from \cite{Fitzpatrick1999} with $R_V=3.1$), taking into account a metallicity effect. We find the relation $M_{W_{JK}}=-2.838 \log P + 0.126\text{[Fe/H]}-1.161$. The scatter around this relation is about 0.05 mag, which is smaller than for Cepheids \citep[0.086 mag; see][]{Breuval2022}

\begin{figure}
   \centering
   \includegraphics[width=\hsize]{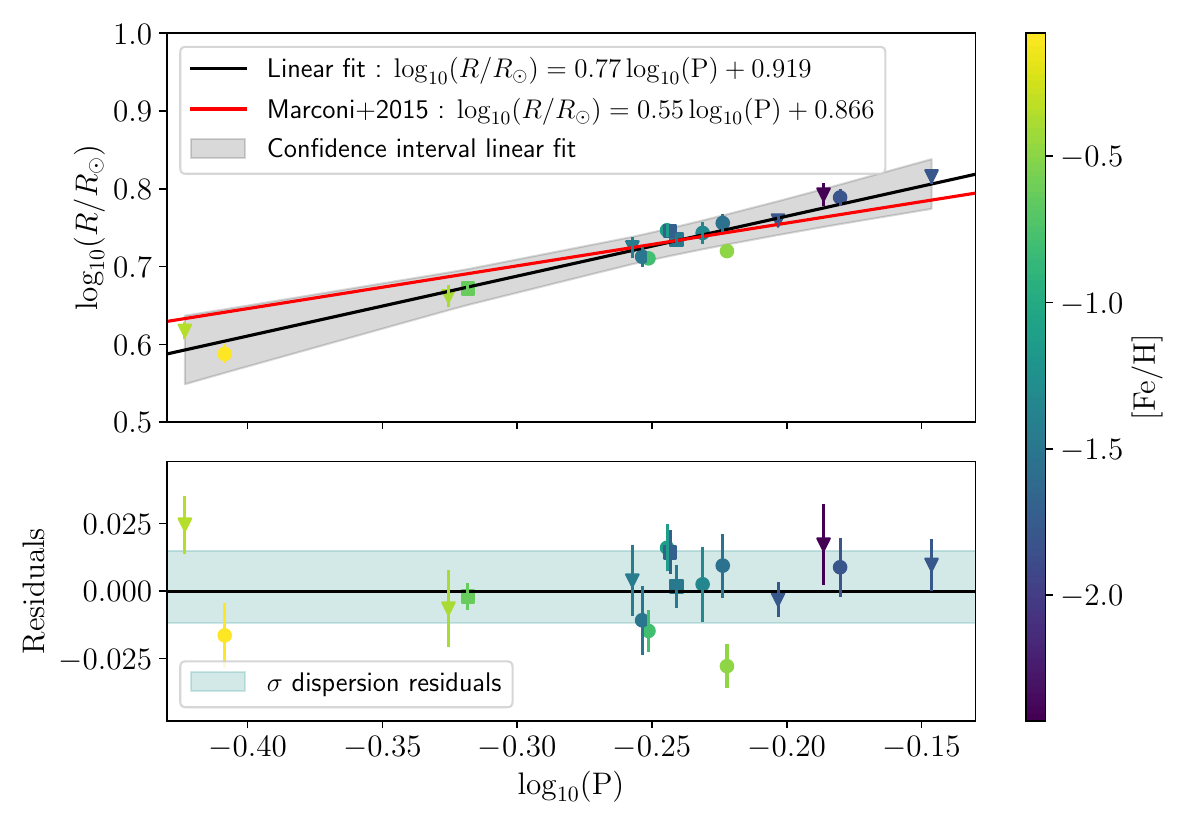}
   \caption{Period--radius relation for the 17 RRLs of our sample. A comparison with theoretical predictions from \cite{Marconi2015} is given in red.}
   \label{PRR}
\end{figure}

Another interesting relation is the period--radius relation, which is more complicated to obtain with the classical observables of a pulsating star, but is easy to reach thanks to SPIPS modelling. This relation is usually tighter than the period--luminosity relation. We show the relation in Fig.~\ref{PRR}, the equation for which is $\log (R/R_\sun)=0.770_{\pm0.003}\log (P)+0.9189_{\pm0.0002}$. We compare with the relation found from non-linear convective models, where 
$\log (R/R_\sun)=0.866_{\pm0.003}+0.55_{\pm0.02}\log (P)$
 \citep{Marconi2015}. This relation is within the confidence interval for our sample. 
 These coefficients are in agreement with those that we find within the range of period containing the majority of our sample.
It is truly remarkable that the period--radius relation is relatively tight. This is a clue that the conversion from photometry to a radius curve is realistic. Hence, the dispersion of the $p$-factors we obtain may come from a spectroscopic effect related to shocks or other specific behaviours of the atmosphere of pulsating stars.

\subsection{HR diagram}

\begin{figure}
   \centering
   \includegraphics[width=\hsize]{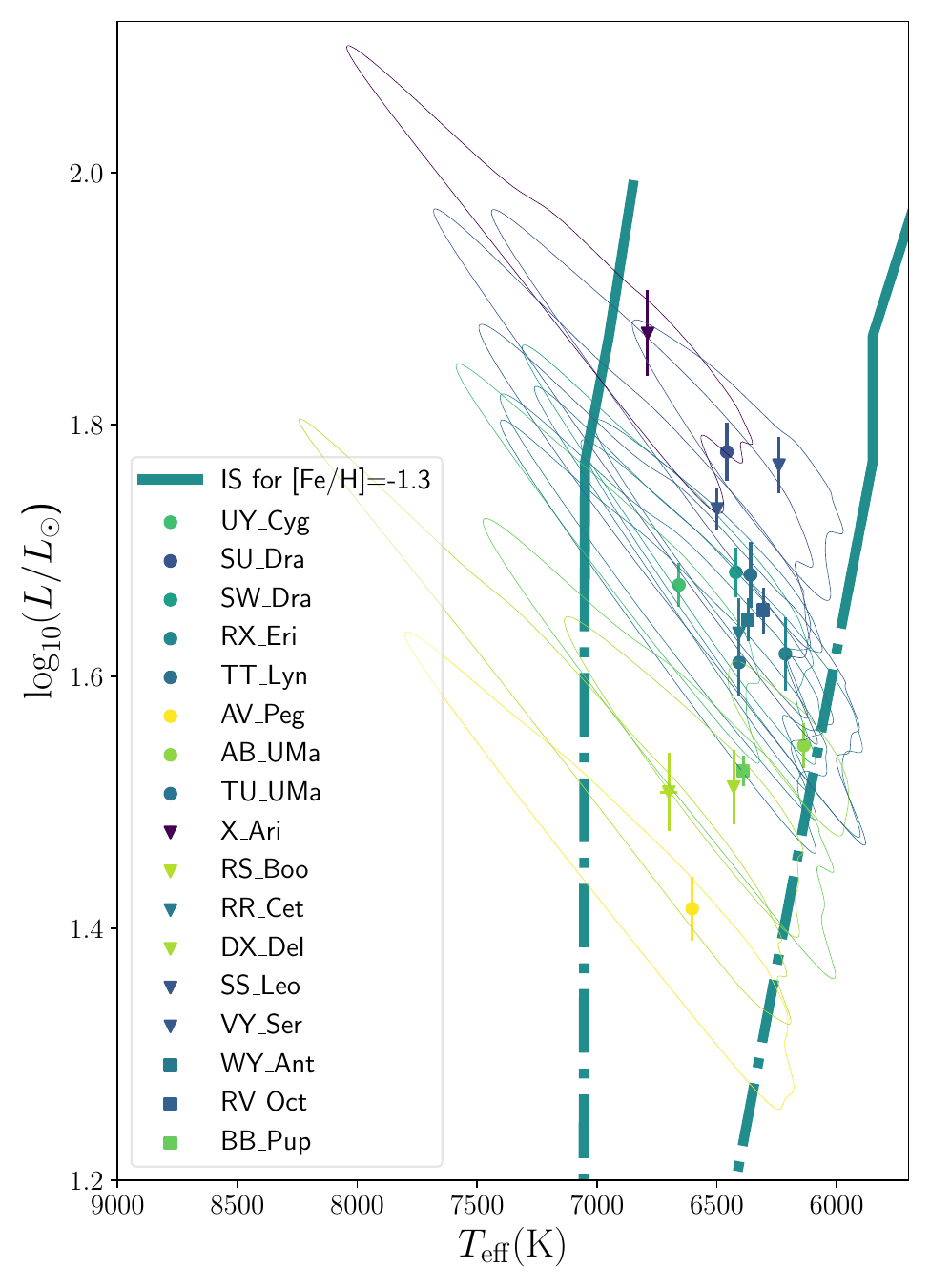}
   \caption{Hertzsprung-Russel diagram showing the pulsation cycle of the 17 modelled RRLs of our sample. Markers indicate the type of RV measurements used in the model (see legend of Fig.~\ref{pP_RRL}). Thick lines are the edges of the instability strip from models \citep{Marconi2015} for a metallicity globally corresponding to RRLs. These predictions have been extended linearly to lower luminosities in semi-dotted lines.}
   \label{HRD}
\end{figure}

Figure~\ref{HRD}  presents the pulsation cycle of all RRLs in the Hertzsprung-Russell (HR) diagram, together with the predicted blue and red edges of the instability strip for a metallicity similar to our RRLs, as predicted by \cite{Marconi2015}.
We note that the RRLs do not usually cool down beyond the red edge. On the contrary, they tend to leave the strip on the blue edge during their pulsation cycle. Nevertheless, their average positions are well confined within the instability strip.

\subsection{Applicability to a \textit{Gaia} sample}

\begin{figure}
   \centering
   \includegraphics[width=\hsize]{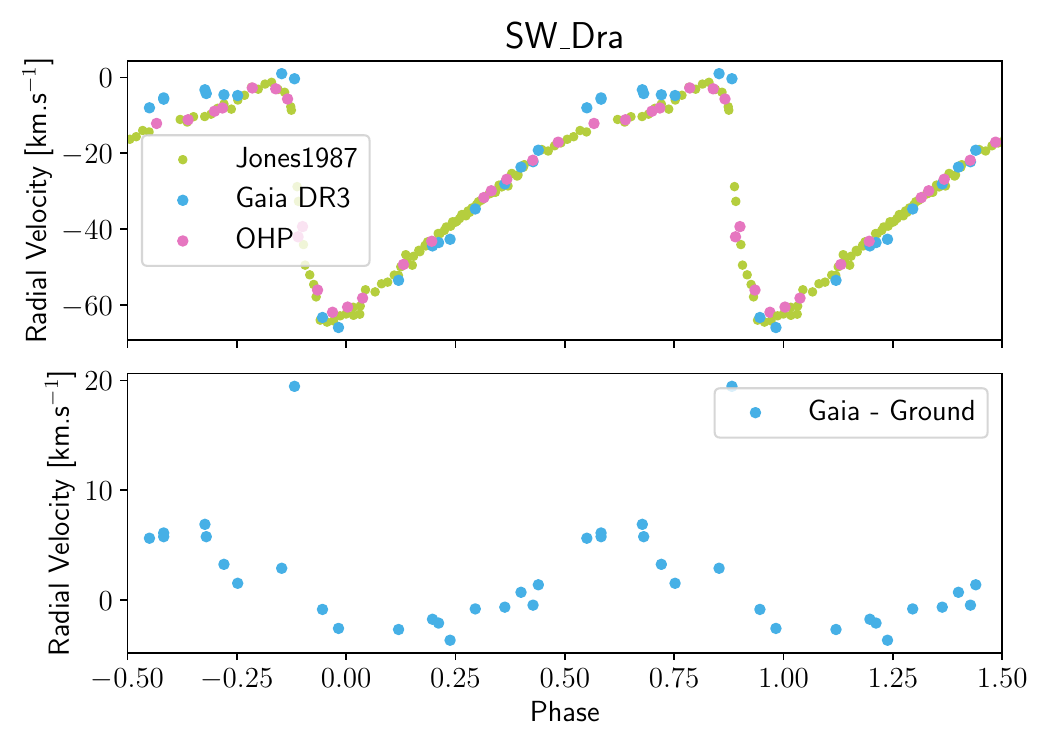}
   \caption{Comparison between the RV curves from the ground \citep[][SOPHIE]{JonesBWSWDra1987} and the \textit{Gaia} series of SW Dra. The bottom plot is the difference between the \textit{Gaia} RV series and the total ground RV curve.}
   \label{RV_comp_Gaia}
\end{figure}

In this paragraph, we aim to test the possibility of performing a precise calibration of the $p$-factor, taking advantage of the \textit{Gaia} photometry and RV curves available in the DR3.
As also reported by \cite{Clementini2023GAIARRL}, we find some significant differences between the RV curves from ground-based spectrographs and from \textit{Gaia}.
For some stars, the \textit{Gaia} Radial Velocity Spectrograph (RVS) curves exhibit a bump before the maximum velocity. We show an example in Fig.~\ref{RV_comp_Gaia}. This difference is likely due to the fact that this measurement is done by cross-correlating a template on a narrow spectrum centred on the calcium triplet.
These spectra can be affected by shock waves in a specific way. Evidence of such shocks mainly comes from studies of Balmer lines \citep{GilletCrowe1988,FokinGillet1997,Gillet2017}, but our knowledge of its effect on the entirety of the spectrum is still incomplete. \cite{Hocde2020b} already showed that a desynchronisation between H$\alpha$ and the calcium triplet lines may occur for long-period Cepheids, for which the surface gravity is lower. The presence of such a bump will increase the amplitude of the integrated RV curve and therefore poses a problem for the PoP method. For the star SW Dra, we estimate that the amplitude of the radius curve will differ by 15\%, and the p-factor will decrease to $p = 1.04$.

\begin{figure*}
\centering
\includegraphics[width=\hsize]{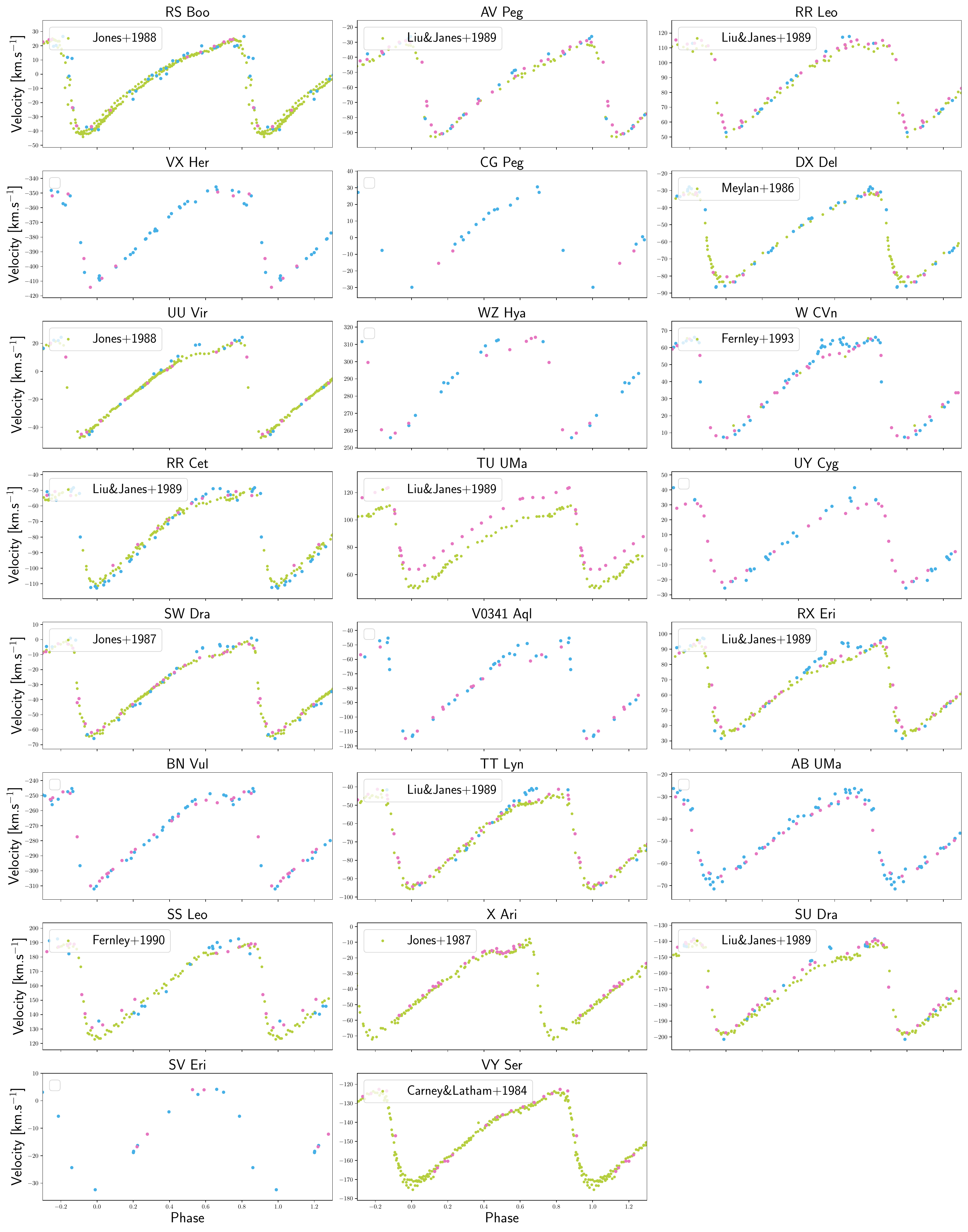}
\caption{The RV curves for the 23 RRLs from measurements with the SOPHIE spectrograph (this work, in pink), data from the references listed in Table \ref{table:1} (in green), and \textit{Gaia} RVS (in blue). We can see the effect of the companion of TU UMa in its RV curve.}
\label{RV_all_RRL_OHP_litt_Gaia}
\end{figure*}

All RV curves from the 23 RRLs in our sample with their \textit{Gaia} RVS and other RV measurements from the literature are plotted in Fig.~\ref{RV_all_RRL_OHP_litt_Gaia}. To enhance readability, only one additional literature dataset is included in the plot. This dataset is either the one used in SPIPS (if available) or the one offering the best phase coverage. At least 12 (V0341 Aql, RR Cet, W CVn, UY Cyg, SW Dra, RX Eri, WZ Hya, RR Leo, SS Leo, TT Lyn, UU Vir and BN Vul) of these stars quite clearly exhibit the same feature as SW Dra. For 6 other stars, we are not able to carry out the comparison, as either \textit{Gaia} or ground-based measurements are missing. Only DX Del appears to really show a similar shape between all curves.

It would be interesting to study the systematic uncertainties linked to the presence of this bump (if it is related to atmospheric parameters of the star or to some range of $p$-factors for instance), how it affects the curve, and why it appears for such spectral lines. This would require \textit{Gaia} data release 4 (DR4), which will provide individual RVS spectra for each observing epoch.

\section{Conclusions}\label{ccl}

In this paper, we present a calibration of the Baade-Wesselink projection factor for a selection of RRLs thanks to new measurements of RVs collected at OHP with the SOPHIE spectrograph.
We computed the $p$-factor for 17 RRLs using \textit{Gaia} DR3 parallaxes.
We examined the influence of the RV measurement method  on $p$  by studying the cross-correlation of our spectra with various masks and templates.
We find that the mask or template for CCF computation does not have a significant impact on $p$ compared to the method employed to derive the RV from the CCF. Our study points out the importance of the precise documentation of how RVs are computed in the literature. However, for consistent RV series, the difference stemming from the methodology used to derive velocities should only shift the projection factor and not affect the scatter.
We calibrated the $p$-factor using the SPIPS modelling tool. As for Cepheids \citep{TrahinpCep}, we see an important scatter in the results, which is greater than the uncertainties in \textit{Gaia} parallaxes. This result has been observed by various authors who calibrated the projection factor, and does not seem to be uniquely correlated with any of the main physical parameters of the stars (period, temperature, metallicity, etc.).
Interestingly, as shown in Fig.~\ref{pP_RRLCep}, the scatter of $p$ seems to be the highest for Cepheids with short periods (below 10 days). However, whether it is a selection effect due to the size of the sample or a real phenomenon is uncertain. We note that we do not find any infrared excess for these RRLs. Although it can have an effect on the $p$-factors of Cepheids \citep{Nardetto2023}, the possible presence of a circumstellar envelope cannot explain the great diversity of the $p$-factors of pulsating stars.

Another remaining uncertainty in the SPIPS analysis is the estimation of the colour excess $E(B-V),$ which we determined from 3D maps of the interstellar medium. Although this value does not have a drastic importance for the $p$-factor, it is still poorly known, because of its degeneracy with the star's effective temperature.
Meanwhile, RRLs in this study are generally nearby, and so the uncertainty is low for this parameter and does not represent a limitation to the present study.

The source of the scatter in the resulting $p$-factors is still poorly understood, but we argue that it is linked to a misunderstanding of the pulsation of the atmosphere of the star. There appears to be a missing ingredient between the pulsational velocity and the motion of line-forming regions, which may be linked to the observations of peculiar features in RVS from \textit{Gaia} DR3.

\begin{acknowledgements}
This work has made use of data from the European Space Agency (ESA) mission {\it Gaia} (\url{http://www.cosmos.esa.int/gaia}), processed by the {\it Gaia} Data Processing and Analysis Consortium (DPAC, \url{http://www.cosmos.esa.int/web/gaia/dpac/consortium}).
Funding for the DPAC has been provided by national institutions, in particular the institutions participating in the {\it Gaia} Multilateral Agreement.
The research leading to these results  has received funding from the European Research Council (ERC) under the European Union's Horizon 2020 research and innovation program (projects CepBin, grant agreement 695099, and UniverScale, grant agreement 951549).
This research has been supported by the Polish-French Marie Skłodowska-Curie and Pierre Curie Science Prize awarded by the Foundation for Polish Science.
A.\,G. aknowledges support from the CONICYT/FONDECYT grant No. 3130361 and ANID-ALMA fund No. ASTRO20-0059. We also acknowledge support from the DIR/WK/2018/09 grant of the Polish Ministry of education and science.
This research has made use of Astropy\footnote{Available at \url{http://www.astropy.org/}}, a community-developed core Python package for Astronomy \citep{Astropy,AstropyProject}, the Numpy library \citep{Numpy}, the Astroquery library \citep{Astroquery} and the Matplotlib graphics environment \citep{Matplotlib}.
We used the SIMBAD and VizieR databases and catalogue access tool at the CDS, Strasbourg (France), and NASA's Astrophysics Data System Bibliographic Services.
The original description of the VizieR service was published in \citep{Vizier}.
This publication makes use of data products from the Two Micron All Sky Survey, which is a joint project of the University of Massachusetts and the Infrared Processing and Analysis Center/California Institute of Technology, funded by the National Aeronautics and Space Administration and the National Science Foundation.
\end{acknowledgements}

\bibliographystyle{aa}
\bibliography{references}

\begin{appendix} %
\section{SPIPS models}\label{appendix:SPIPS}
\begin{figure*}
        \centering
        \includegraphics[width=\hsize]{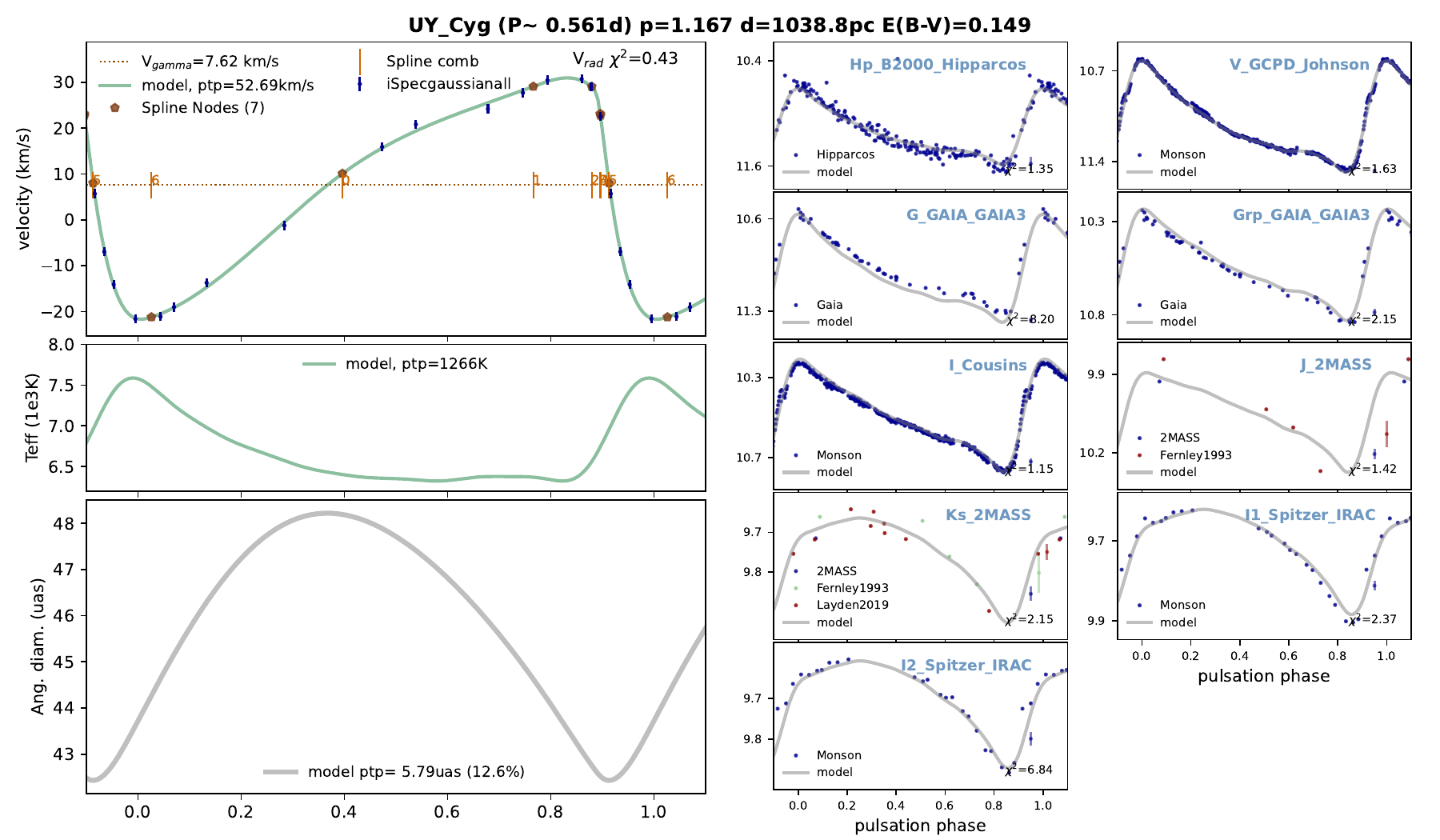}
        \caption{SPIPS model of UY Cyg with RV from SOPHIE spectra (Gaussian fit of the CCF using the binary mask built for RRLs with all unblended lines). Photometry measurements are from \cite{MonsonPhotRRL}, \cite{FernleyPhotRV1993}, and \cite{Layden2019} (70\% of available photometric measurements).}
\end{figure*}

\begin{figure*}
        \centering
        \includegraphics[width=\hsize]{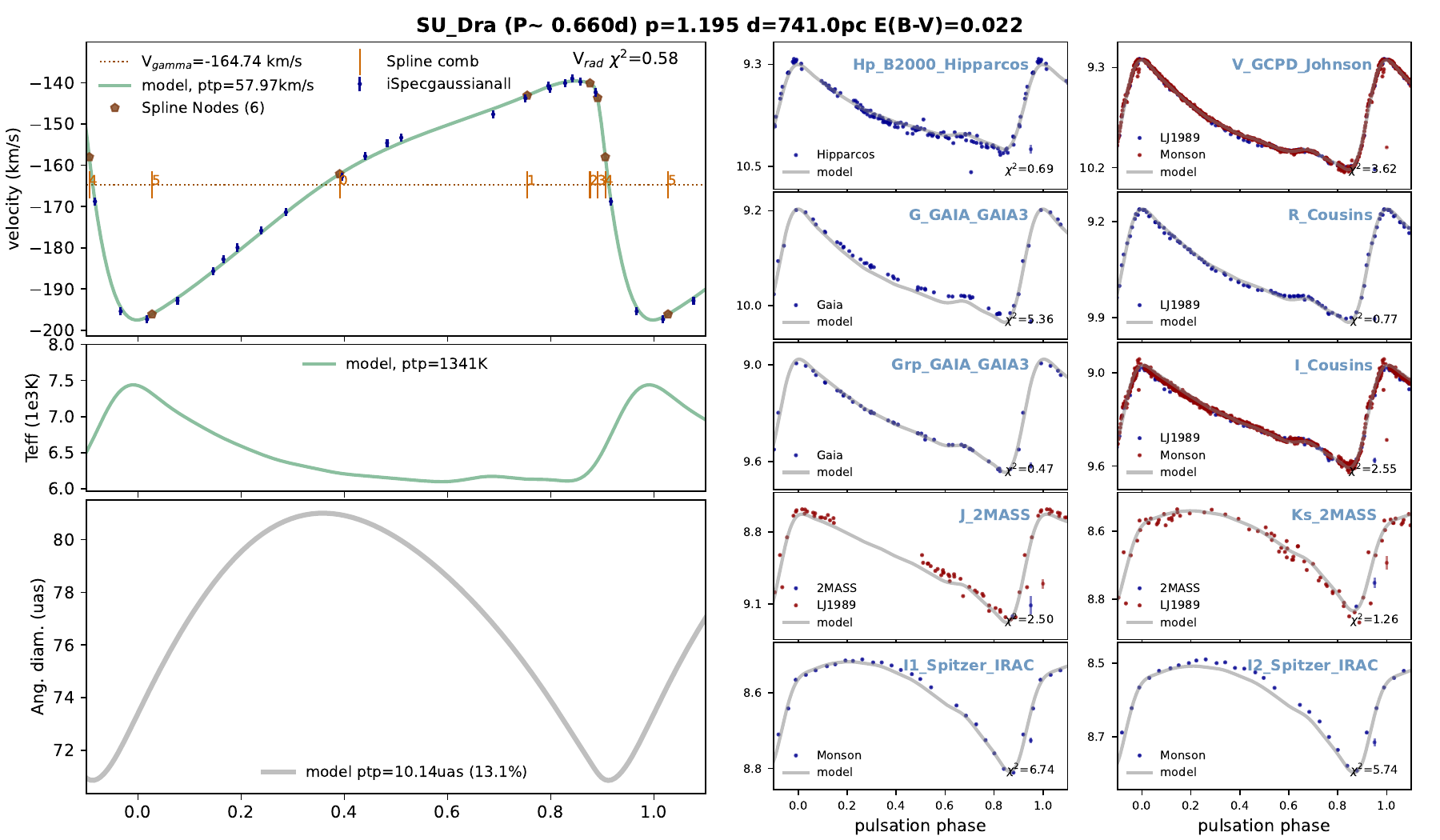}
        \caption{SPIPS model of SU Dra with RV from SOPHIE spectra (Gaussian fit of the CCF using the binary mask built for RRLs with all unblended lines). Photometry measurements are from \cite{MonsonPhotRRL} and \cite{LJBW1989} (70\% of available photometric measurements).}
\end{figure*}

\begin{figure*}
        \centering
        \includegraphics[width=\hsize]{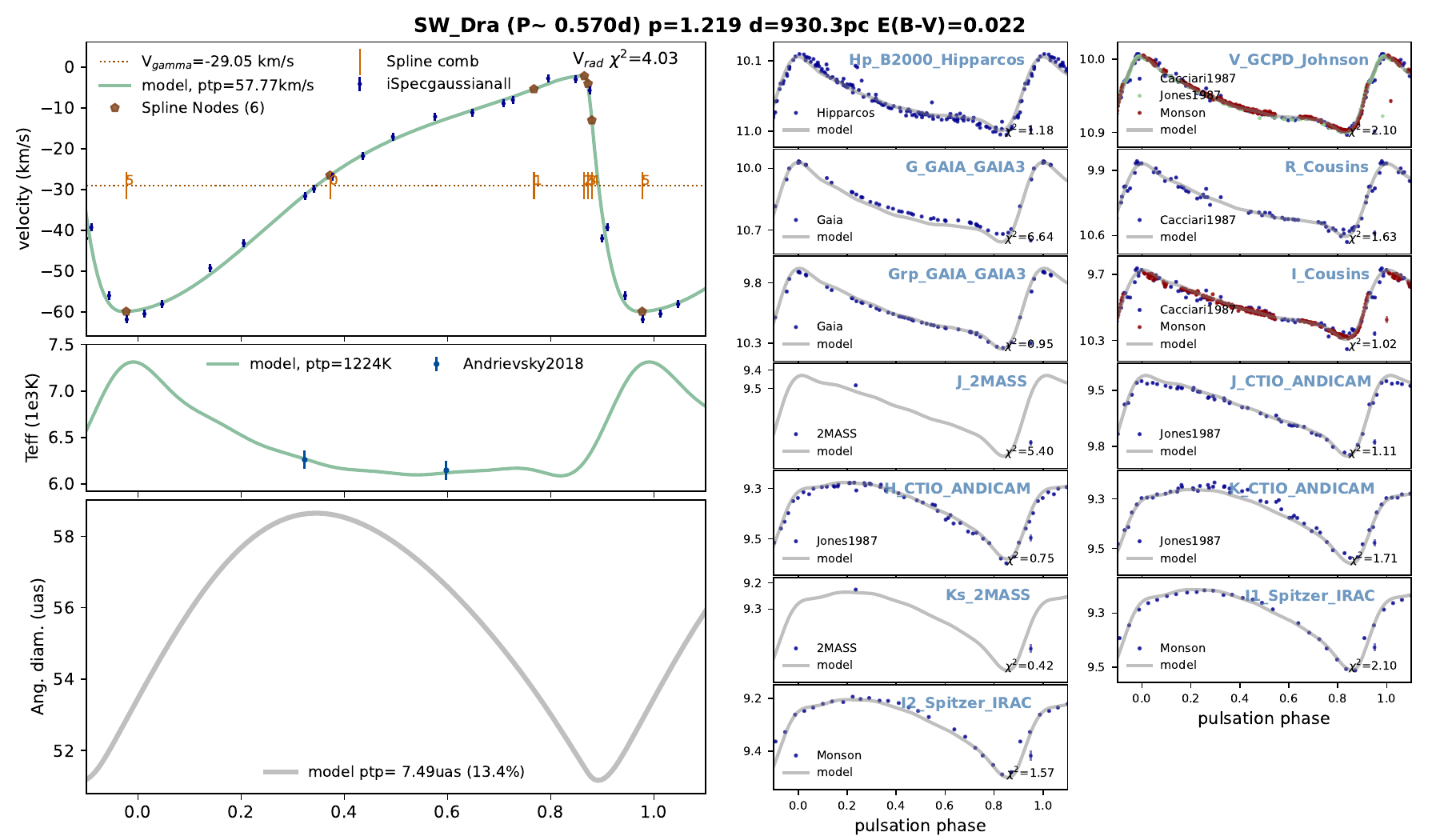}
        \caption{SPIPS model of SW Dra with RV from SOPHIE spectra (Gaussian fit of the CCF using the binary mask built for RRLs with all unblended lines). Photometry measurements are from \cite{MonsonPhotRRL}, \cite{CacciariBW1987}, and \cite{JonesBWSWDra1987} (68\% of available photometric measurements).}
\end{figure*}

\begin{figure*}
        \centering
        \includegraphics[width=\hsize]{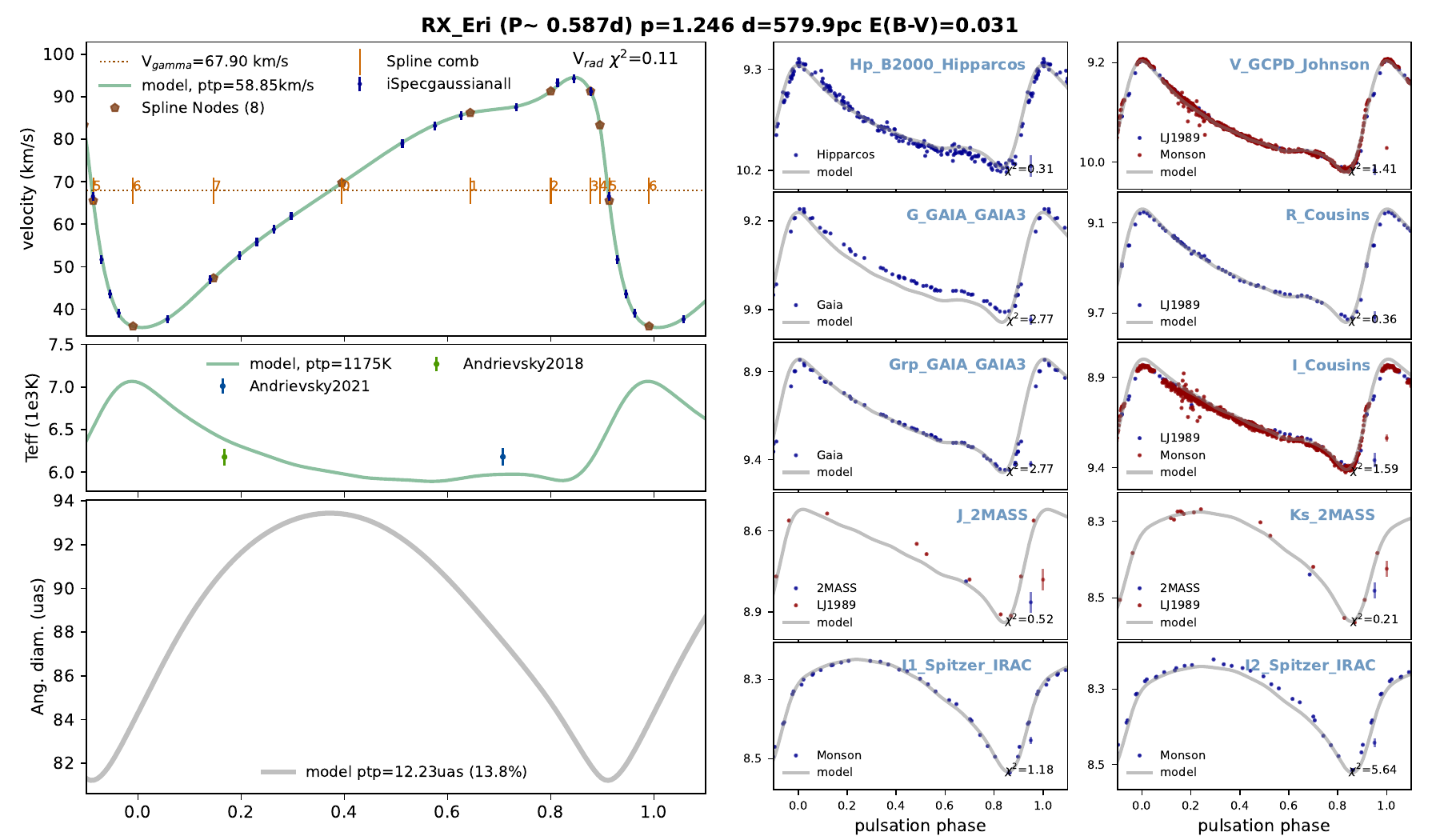}
        \caption{SPIPS model of RX Eri with RV from SOPHIE spectra (Gaussian fit of the CCF using the binary mask built for RRLs with all unblended lines). Photometry measurements are from \cite{MonsonPhotRRL} and \cite{LJBW1989} (64\% of available photometric measurements).}
\end{figure*}

\begin{figure*}
        \centering
        \includegraphics[width=\hsize]{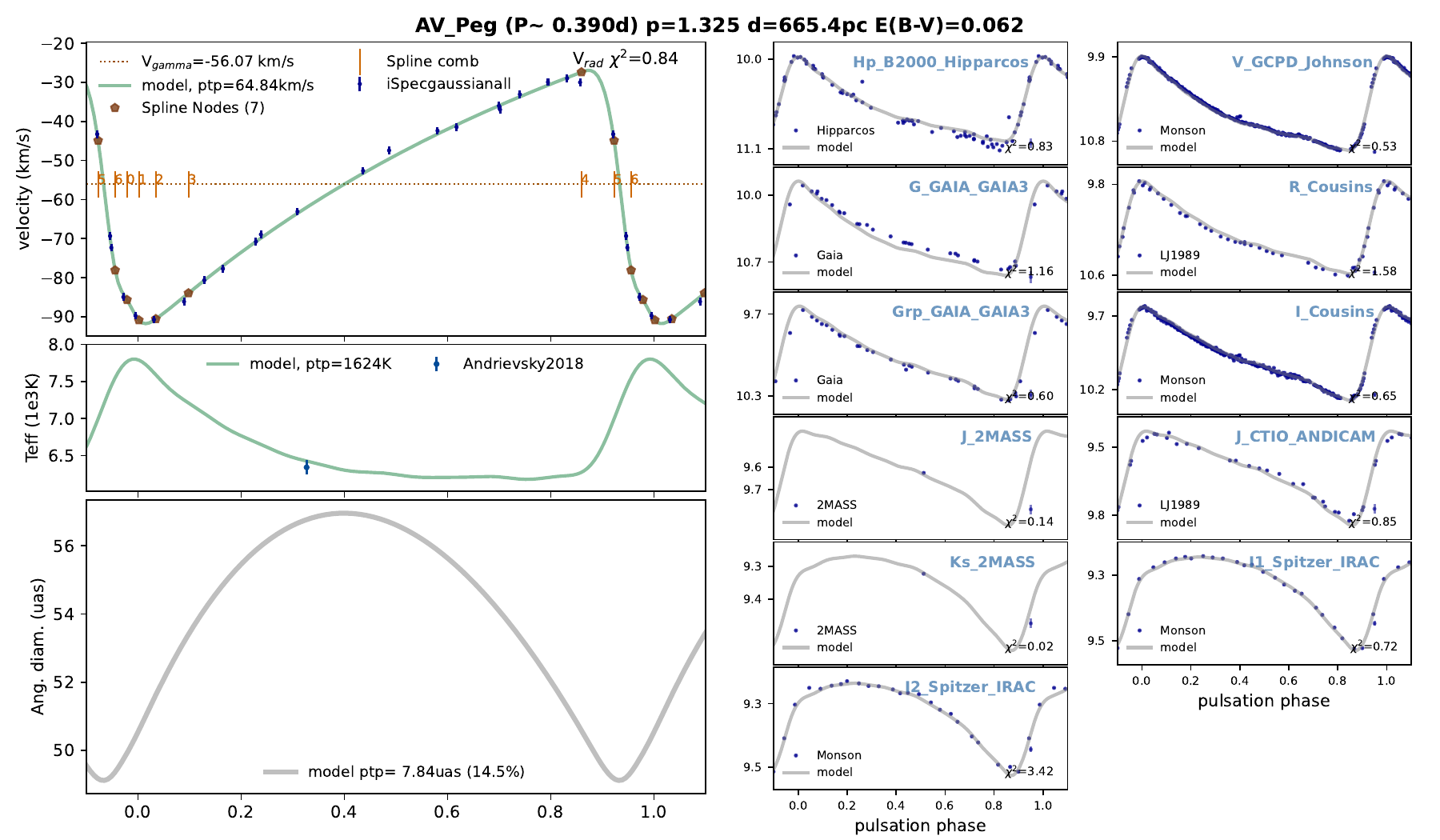}
        \caption{SPIPS model of AV Peg with RV from SOPHIE spectra (Gaussian fit of the CCF using the binary mask built for RRLs with all unblended lines). Photometry measurements are from \cite{MonsonPhotRRL} and \cite{LJBW1989} (56\% of available photometric measurements).}
\end{figure*}

\begin{figure*}
        \centering
        \includegraphics[width=\hsize]{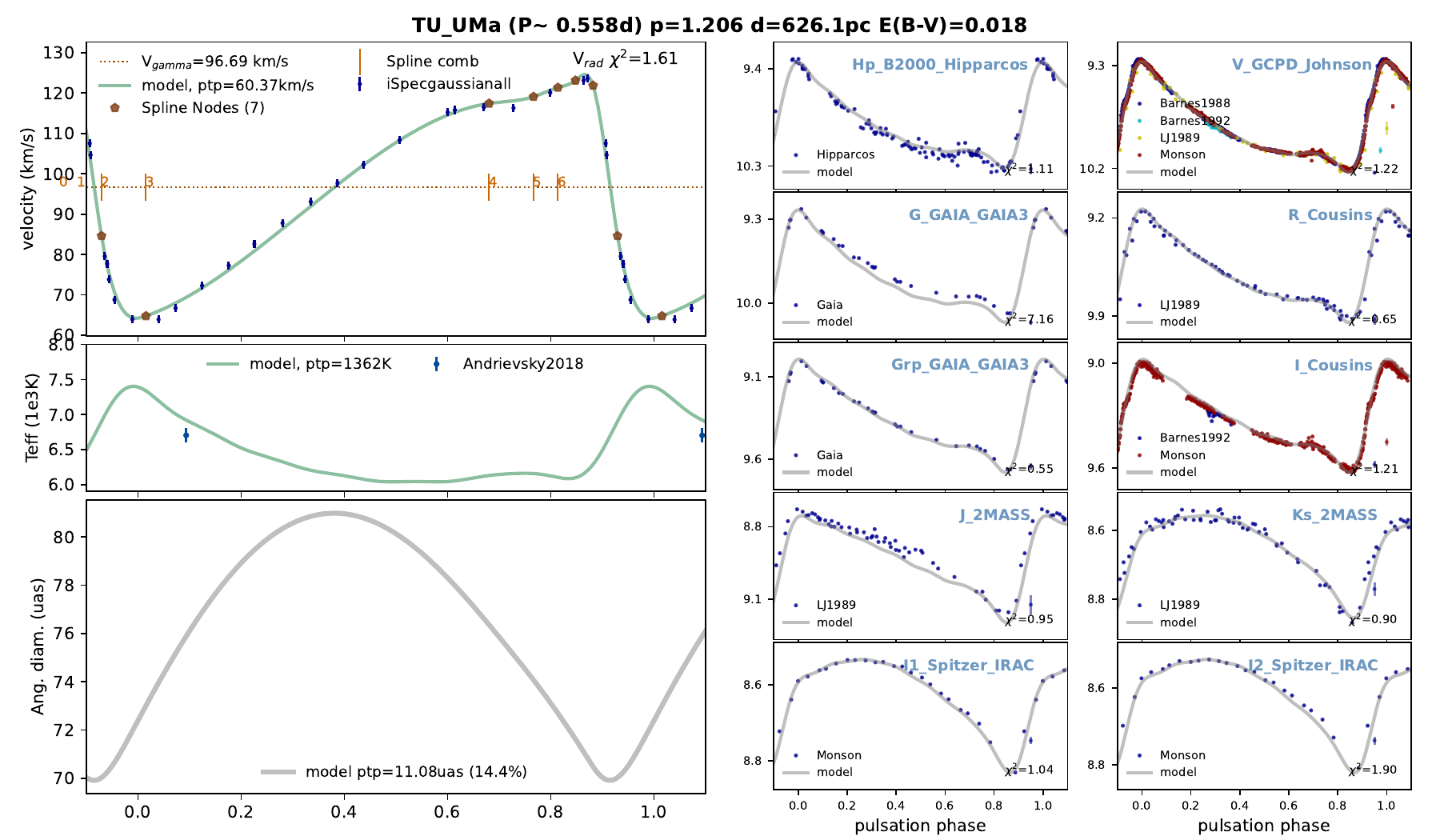}
        \caption{SPIPS model of TU UMa with RV from SOPHIE spectra (Gaussian fit of the CCF using the binary mask built for RRLs with all unblended lines). Photometry measurements are from \cite{MonsonPhotRRL}, \cite{LJBW1989}, \cite{BarnesBW1988}, and \cite{BarnesPhot1992} (42\% of available photometric measurements).}
\end{figure*}

\begin{figure*}
        \centering
        \includegraphics[width=\hsize]{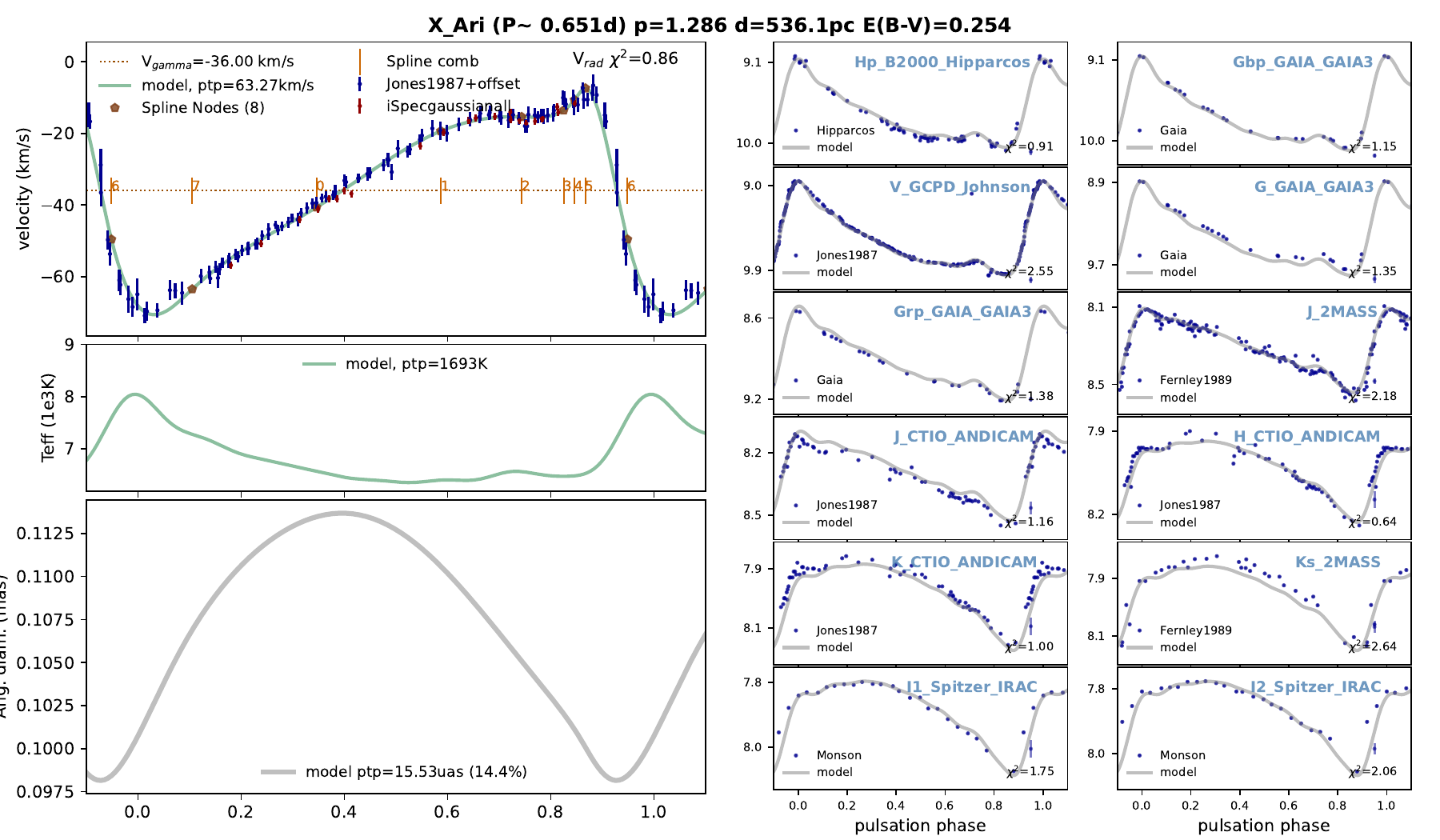}
        \caption{SPIPS model of X Ari with RV from \cite{JonesBWXAri1987} scaled to RV measurements (Gaussian fit of the CCF using the binary mask built for RRLs with all unblended lines), offset=1.285013 km.s$^{-1}$. Photometry measurements are from \cite{MonsonPhotRRL}, \cite{JonesBWXAri1987}, and \cite{FernleyBWXAri1989} (23\% of available photometric measurements).}
\end{figure*}

\begin{figure*}
        \centering
        \includegraphics[width=\hsize]{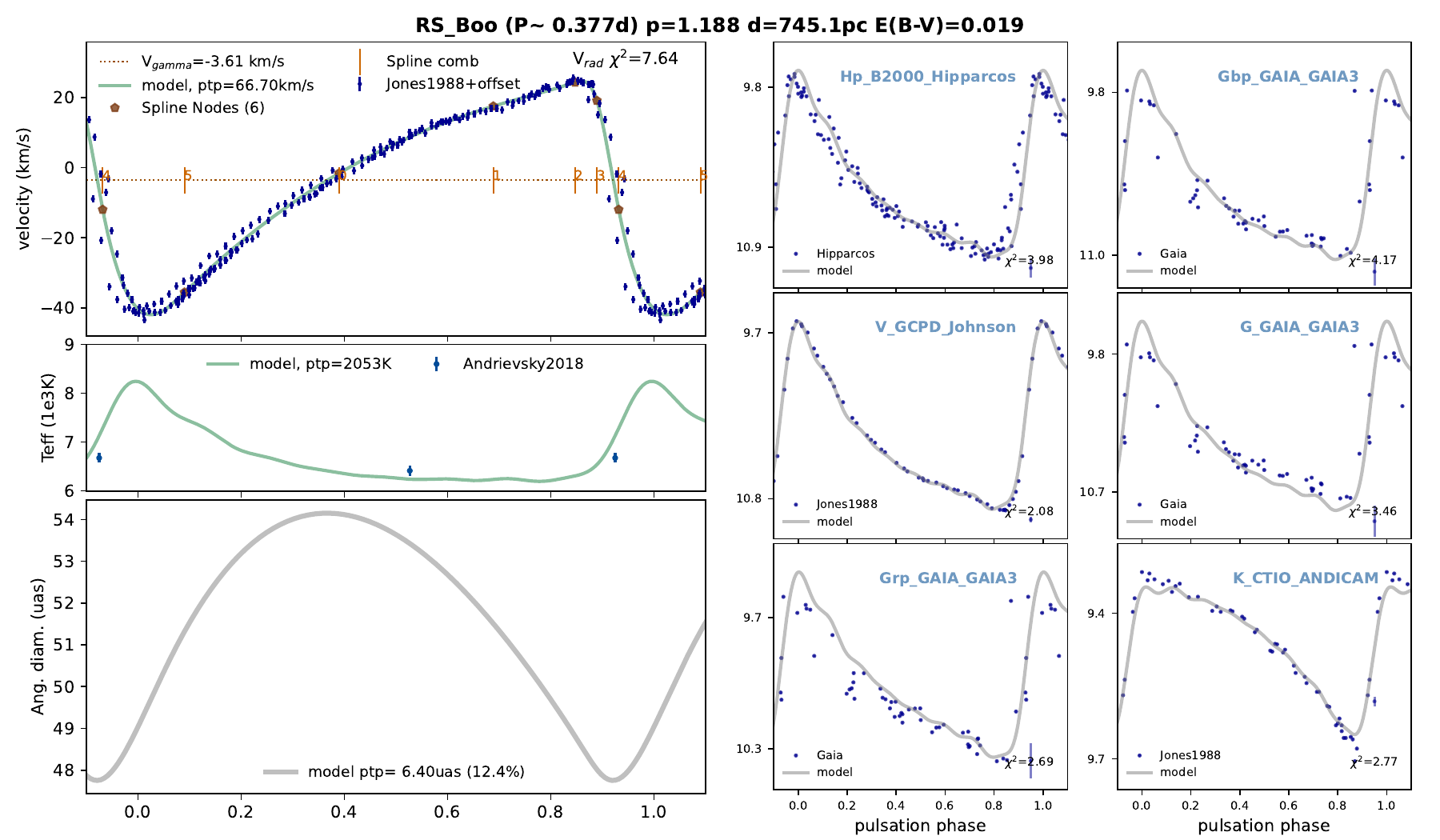}
        \caption{SPIPS model of RS Boo with RV from \cite{JonesBW1988} scaled to RV measurements (Gaussian fit of the CCF using the binary mask built for RRLs with all unblended lines), offset=0.64241 km.s$^{-1}$. Photometry measurements are from \cite{JonesBW1988} (87\% of available photometric measurements). 
 }
\end{figure*}

\begin{figure*}
        \centering
        \includegraphics[width=\hsize]{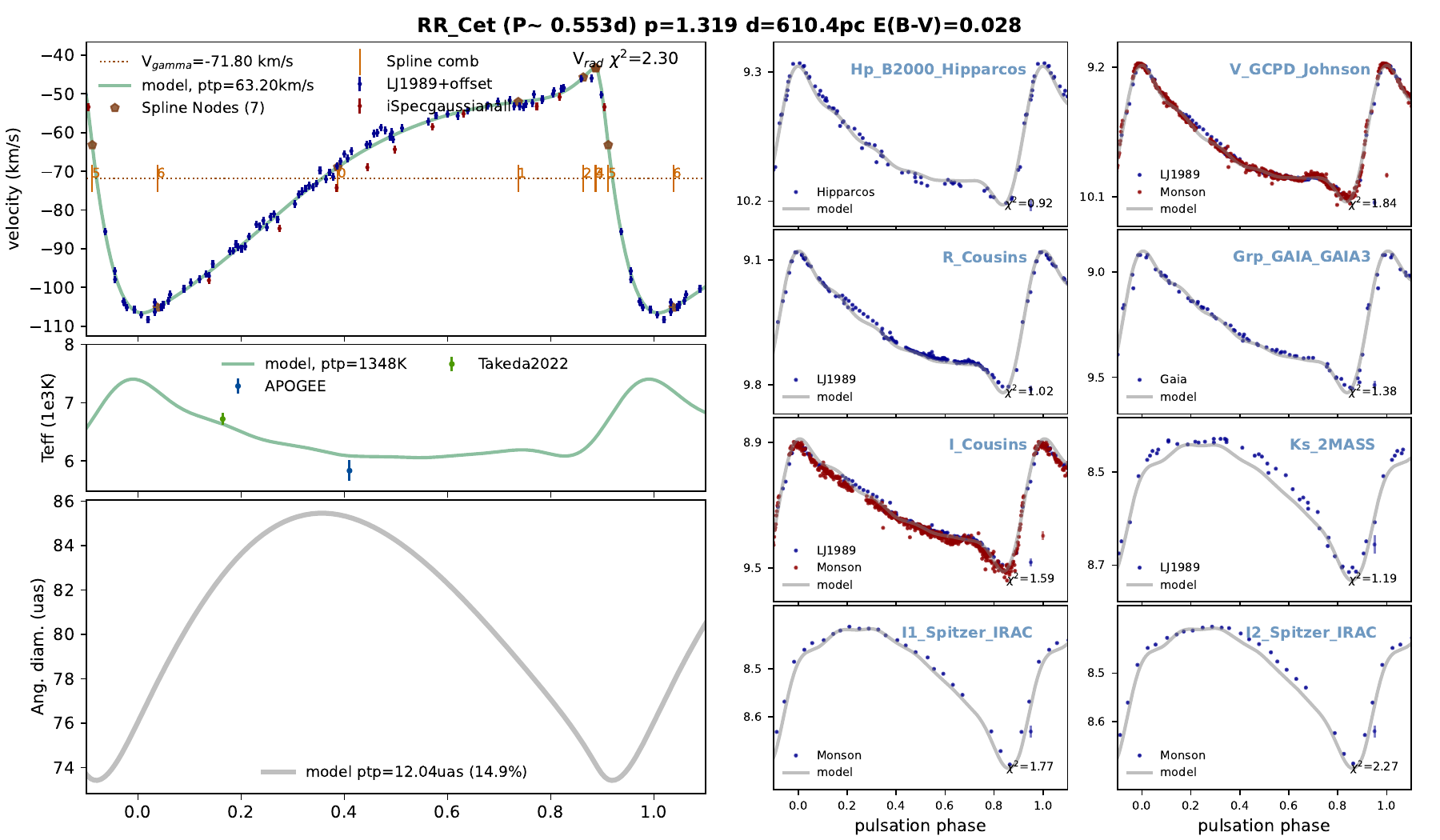}
        \caption{SPIPS model of RR Cet with RV from \cite{LJBW1989} scaled to RV measurements (Gaussian fit of the CCF using the binary mask built for RRLs with all unblended lines), offset=2.952771 km.s$^{-1}$. Photometry measurements are from \cite{MonsonPhotRRL} and \cite{LJBW1989} (21\% of available photometric measurements).}
\end{figure*}

\begin{figure*}
        \centering
        \includegraphics[width=\hsize]{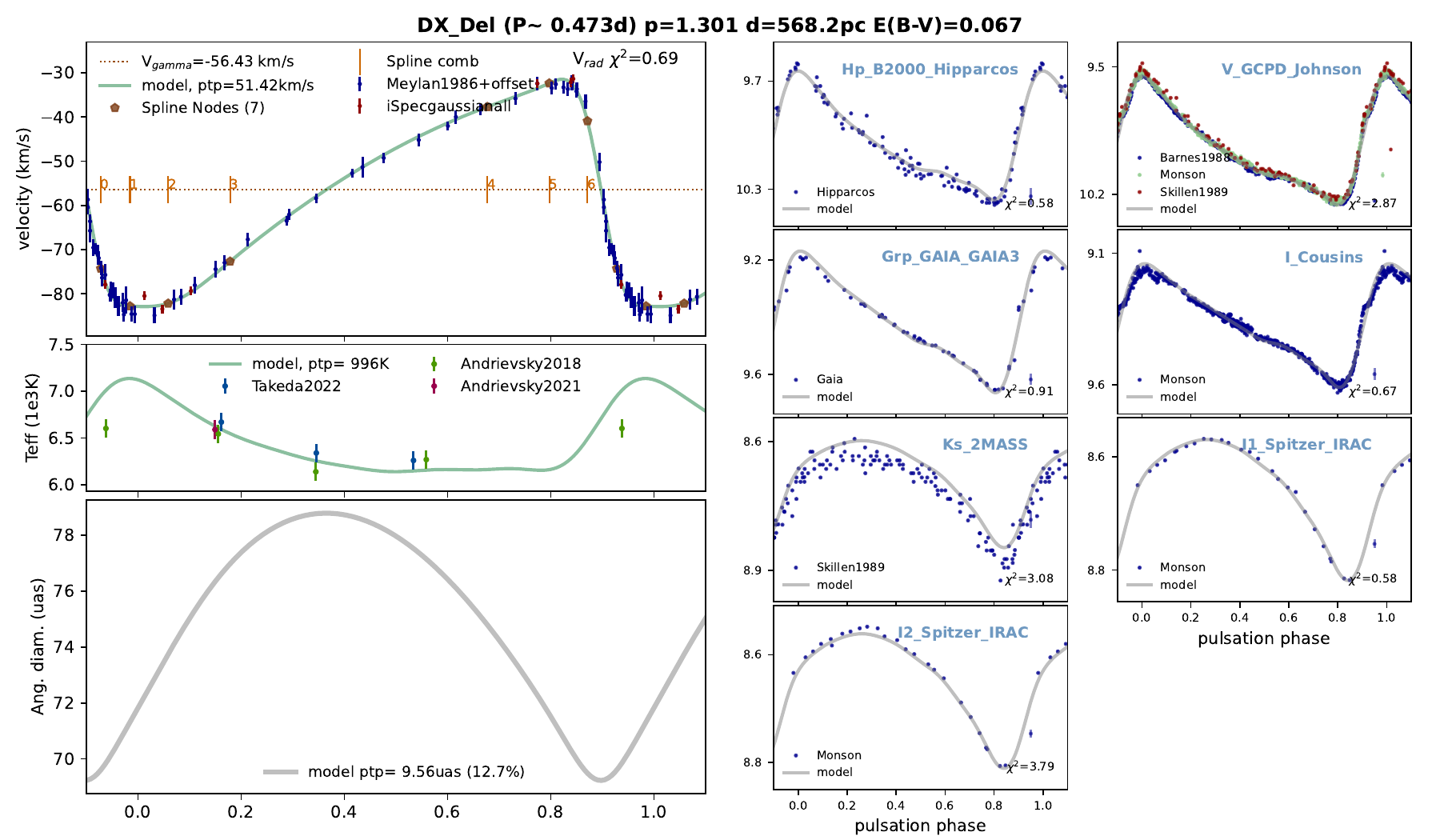}
        \caption{SPIPS model of DX Del with RV from \cite{MeylanBW1986} scaled to RV measurements (Gaussian fit of the CCF using the binary mask built for RRLs with all unblended lines), offset=-1.173494 km.s$^{-1}$. Photometry measurements are from \cite{MonsonPhotRRL}, \cite{BarnesBW1988}, and \cite{SkillenBW1993} (27\% of available photometric measurements).}
\end{figure*}

\begin{figure*}
        \centering
        \includegraphics[width=\hsize]{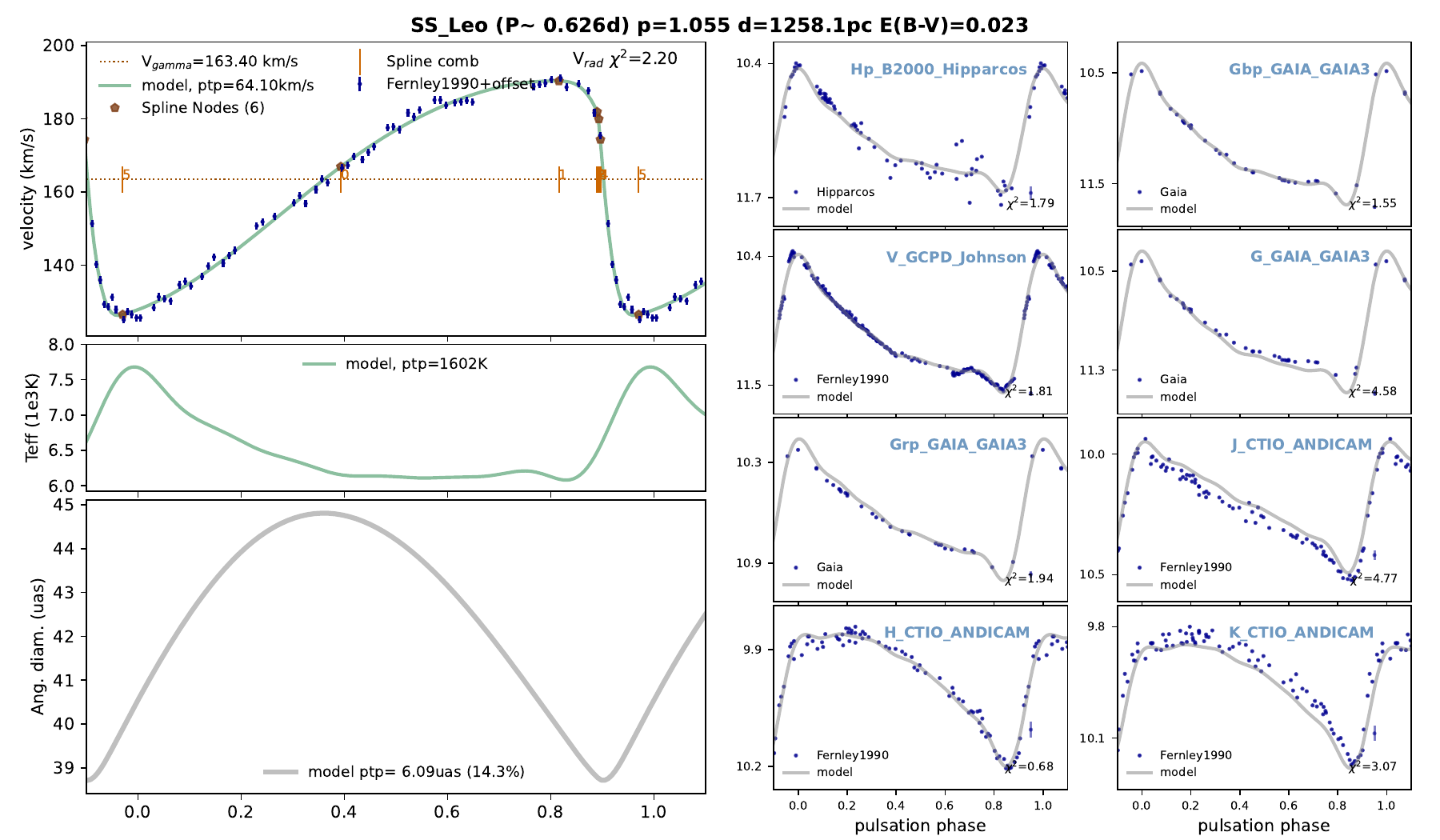}
        \caption{SPIPS model of SS Leo with RV from \cite{FernleySSLeoVYSer1990} scaled to RV measurements (Gaussian fit of the CCF using the binary mask built for RRLs with all unblended lines), offset=2.219878 km.s$^{-1}$. Photometry measurements are from \cite{FernleySSLeoVYSer1990} (44\% of available photometric measurements).}
\end{figure*}

\begin{figure*}
        \centering
        \includegraphics[width=\hsize]{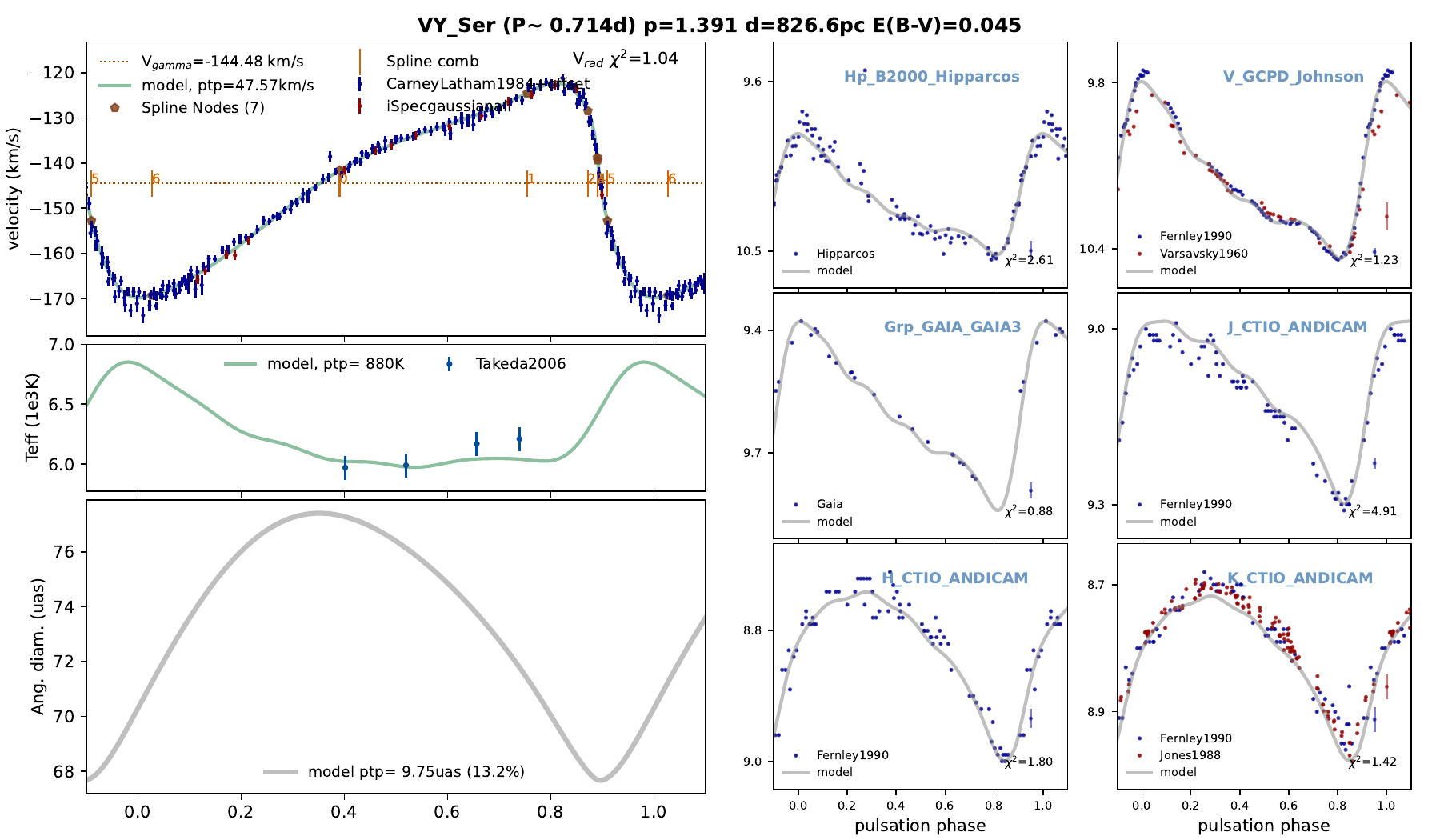}
        \caption{SPIPS model of VY Ser with RV from \cite{CLBWVYSer1984} scaled to RV measurements (Gaussian fit of the CCF using the binary mask built for RRLs with all unblended lines), offset=1.658486 km.s$^{-1}$. Photometry measurements are from \cite{FernleySSLeoVYSer1990} and \cite{VarsavskyPhotRV1960} (41\% of available photometric measurements). 
 }
\end{figure*}

\begin{figure*}
        \centering
        \includegraphics[width=\hsize]{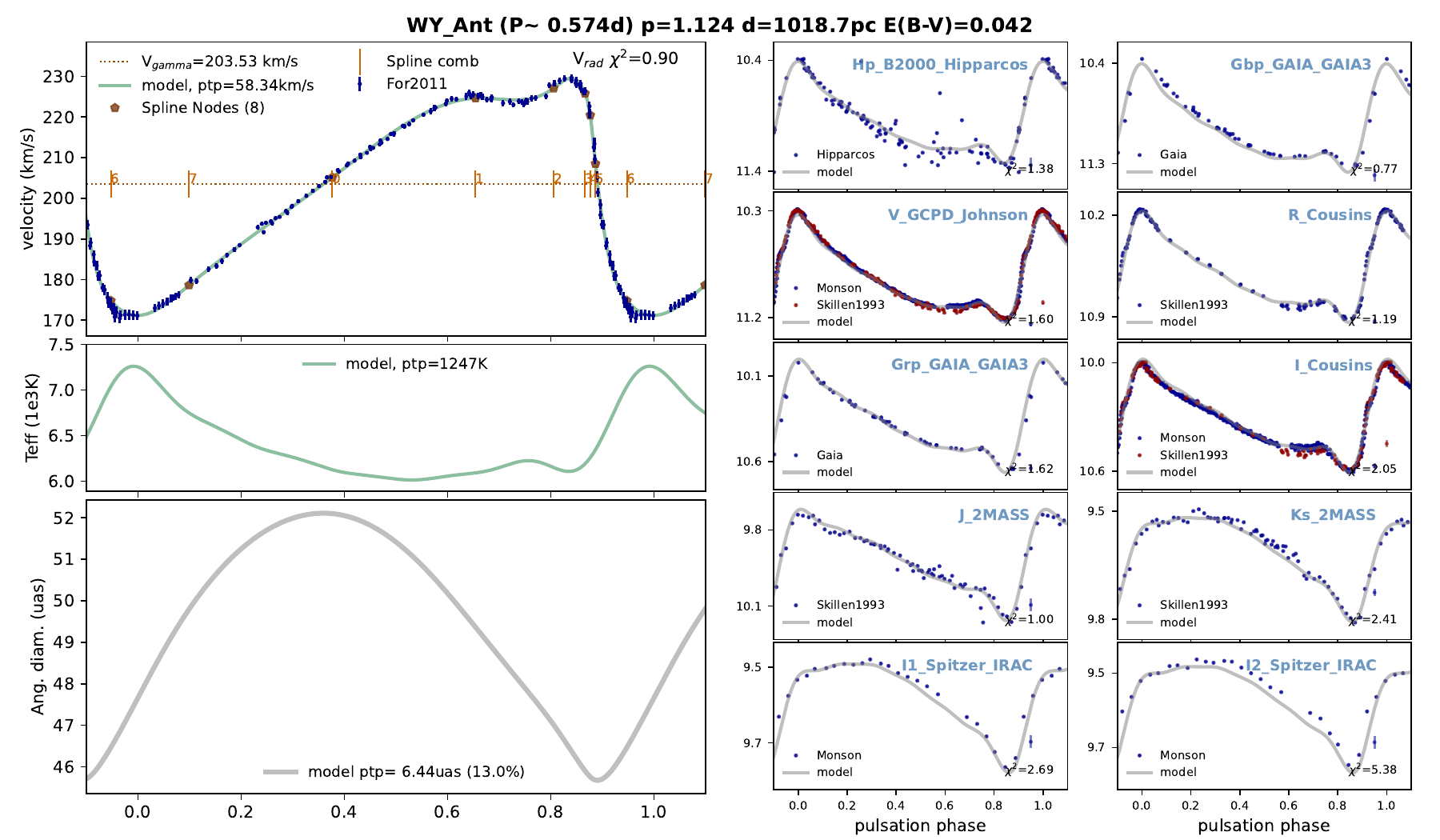}
        \caption{SPIPS model of WY Ant with RV from \cite{ForRV2011} (not observed at OHP). Photometry measurements are from \cite{MonsonPhotRRL} and \cite{SkillenBW1993} (66\% of available photometric measurements).}
\end{figure*}

\begin{figure*}
        \centering
        \includegraphics[width=\hsize]{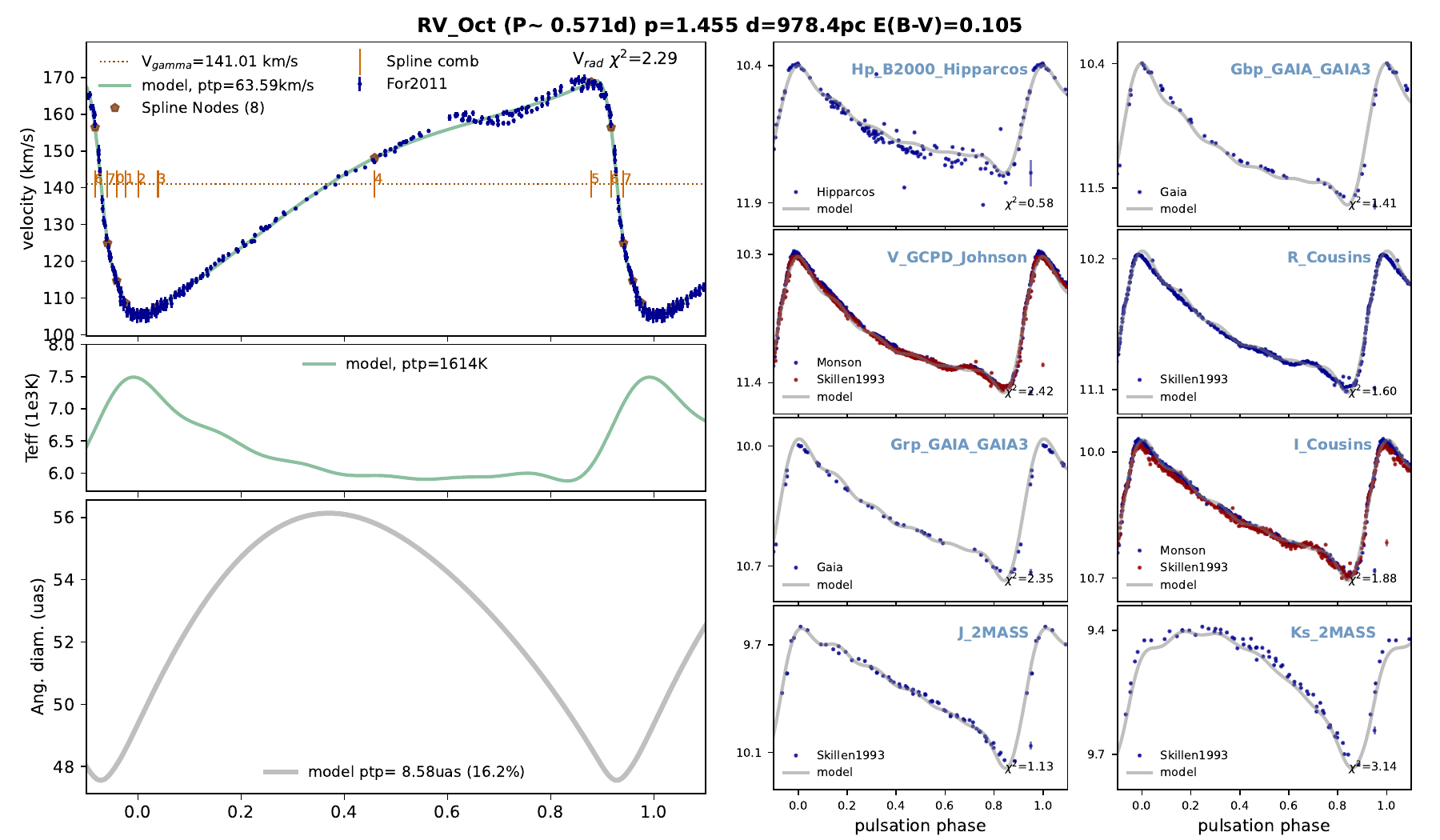}
        \caption{SPIPS model of RV Oct with RV from \cite{ForRV2011} (not observed at OHP). Photometry measurements are from \cite{MonsonPhotRRL} and \cite{SkillenBW1993} (80\% of available photometric measurements).}
\end{figure*}

\begin{figure*}
        \centering
        \includegraphics[width=\hsize]{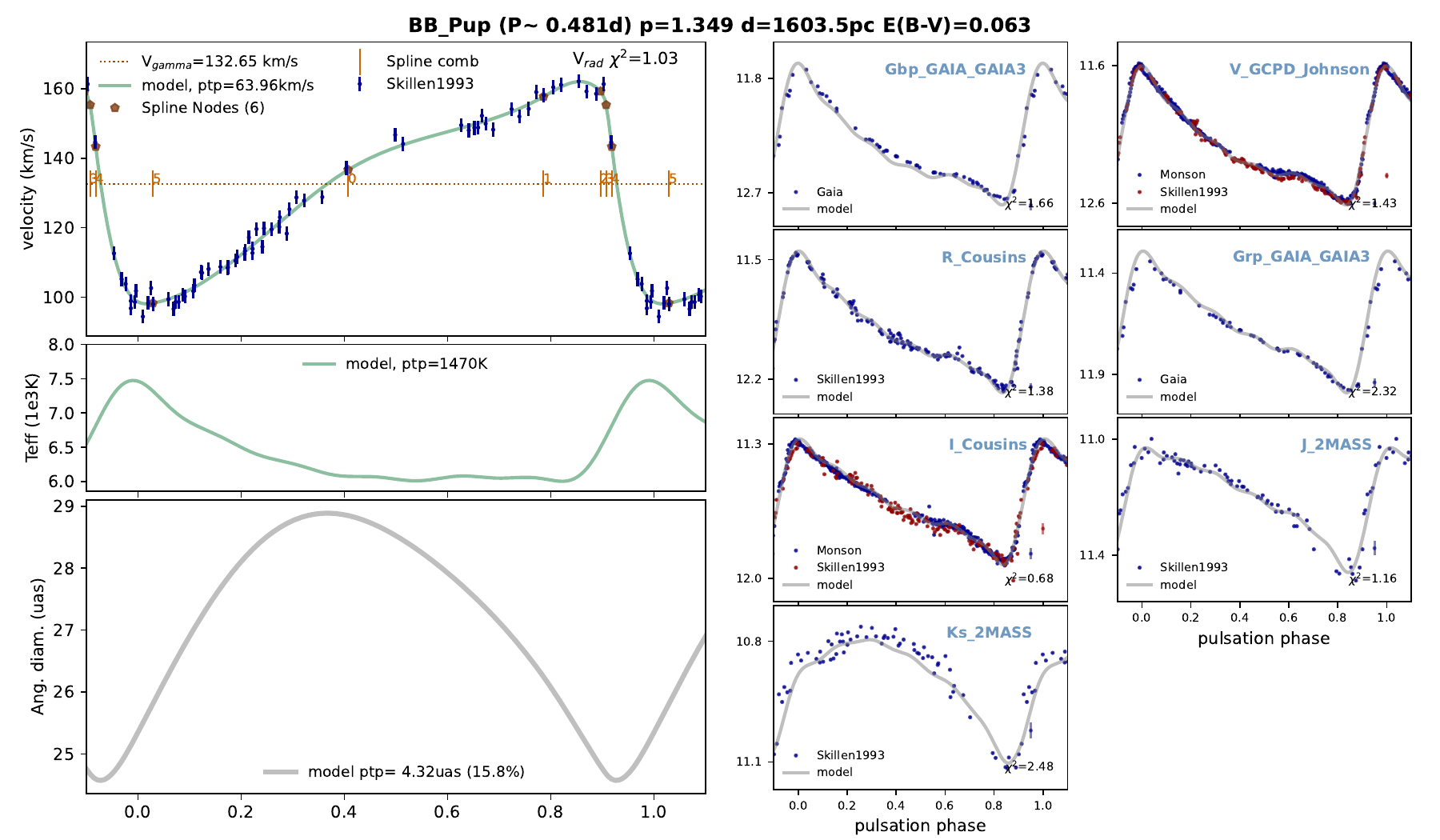}
        \caption{SPIPS model of BB Pup with RV from \cite{SkillenBW1993} (not observed at OHP). Photometry measurements are from \cite{MonsonPhotRRL} and \cite{SkillenBW1993} (64\% of available photometric measurements).}
\end{figure*}

\end{appendix}

\end{document}